\documentstyle[a4wide,newapax,apalikex,graphicx]{article}
\let\myund\&
\renewcommand{\&}{and}

\begin{document}
\title{A new binary decagonal Frank-Kasper quasicrystal phase}
\author{Johannes~Roth$^{\mbox{\small$\ast$}}$\\ {\normalsize Institut f\"ur
    Theoretische und Angewandte Physik, Universit\"at Stuttgart,} \\ 
  {\normalsize Pfaffenwaldring 57, D-70550 Stuttgart, Germany}\\[1ex]
  Christopher~L.~Henley\\ {\normalsize $^{\mbox{\small$\ast$}}$Laboratory of
    Atomic and Solid-State Physics, Cornell University,}\\ {\normalsize
    Ithaca, NY 14853--2501, USA}}

\maketitle

\begin{abstract}
A structure model of atoms of two sizes, interacting with Lennard-Jones
potentials and simulated by molecular dynamics, was observed to freeze into a
decagonal quasicrystal dominated by Frank-Kasper coordination shells and
closely related to the Henley-Elser model for icosahedral quasicrystals. 
Idealized structure models can be described as decorations of
triangle-rectangle and rhombus tilings. Equilibrium properties of the
idealized model have been determined by molecular dynamics simulations
and a high stability of the model and a low jump rate of the atoms have also
been observed.\\


preprint ITAP-3/96-1

\end{abstract}

\section{ Introduction}

The well-studied quasicrystals fall essentially into two families, 
the Al-transition metal family exemplified by i(AlPdMn)
and the Frank-Kasper family \cite{FK1,FK2,shoe57,shoe87} exemplified by
i(AlCuLi). Decagonal quasicrystals are known experimentally only in the first
family, except for one claim \cite{chinese}.

Recently, three dimensional model systems have been found which,
in simulations, freeze into a quasicrystal phase with 
Frank-Kasper local structure:
a monatomic  system with a special potential 
freezes into a dodecagonal quasicrystal \cite{dzug}, and
a system of large (L) and small (S) atoms with Lennard-Jones
interactions freezes into an icosahedral phase \cite{rotcool}, 
in fact a simplified version of the Henley-Elser structure model for
i(AlCuLi)  and i(AlZnMg) \cite{henels}.
Since Frank-Kasper systems are  dominated by $sp$ conduction electrons,
it is believed that simple pair potentials (as used in these
simulations)
can offer a good approximation to the structure energy \cite{haf}.
The L and S atoms used by Roth et al.\ \citeyear{rotcool} had single-well 
potentials where the ideal distances 
are non-additive $(r_{LL} \approx r_{LS} \approx 1.15 r_{SS})$.

This paper reports on a new {\it decagonal} Frank-Kasper phase 
which froze from the liquid in molecular dynamics simulations
of a model system using practically the same potentials as 
Roth et al.\ \citeyear{rotcool} (see Sec.~\ref{simulations}). 
No realization of this structure in nature has been established
(except conceivably in d(FeNb), as discussed in Sec.~\ref{comparison}).
However, in the absence of a {\it realistic} model system, a toy model
system with an equilibrium quasicrystal phase can be quite useful.
(To date, microscopically derived potentials for particular alloy systems have 
not been used in simulations of freezing from the melt \cite{mihII}, but
only for comparisons of trial structures at zero temperature, in 
the Al-transition metal case (Phillips, Deng, Carlsson and Murray, 1991; Zou
and Carlsson, 1994; Widom, Phillips, Zou and Carlsson, 1995; Widom and
Phillips, 1993; Phillips and Widom, 1994; Mihalcovi{\v c}, Zhu, Henley and
Oxborrow, 1996; Mihalcovi{\v c} et al.\ 1996b) 
\nocite {carlsson1,carlsson2,carlsson3,wid92,phil,mihI,mihII}).
Of course, there is some fundamental interest in the existence
of any microscopic model of interacting atoms
which has a quasicrystal equilibrium state.
Furthermore, within such a model
one can study the phonon/phason coupling, 
the energy changes and barriers corresponding to tile flips,  
the phason elastic constants with their temperature dependence, 
and can derive from microscopics 
an effective Hamiltonian in terms of tile-tile interactions (Mihalcovi{\v c} et
al.\ 1996a). 
In two dimensions, the ``binary tiling'' case \cite{lan86,wid87}
has served this purpose, 
but it has quite unrealistic bond radii, in contrast to
our model potentials 
(which were in fact tailored to favor the i(AlCuLi) structure \cite{rotstab}).

The bulk of our paper
presents an idealized decoration model which  was constructed by
abstracting the patterns observed in our simulations.
The model has several variants, corresponding to closely related
atomic structures, yet described by different tilings
(See Sec.~\ref{dsdm}).
Thus it exemplifies a system where an effective tile-tile Hamiltonian 
may select a less disordered subensemble of a random tiling ensemble
\cite{jeong}.  
We have constructed the acceptance domains of a quasiperiodic
version of our model (Sec.~\ref{environments}). 
Finally, by applying information from the ideal model, 
we adjusted the potentials
so as to obtain a better-ordered structure after quenching (Sec.~\ref{XXX}).
In the concluding section, 
we have discussed how our model is related to others, and how 
realistically it extends beyond our small system size.

\section { Initial Simulations} \label{simulations}

Our decagonal phase was first seen in a slow-cooling simulation of
a binary liquid. (Further details about this simulation and other structures
observed, as well as the motivation for our initial choice of parameters for
the potential, can be found (Roth et al.\ 1995). \nocite{rotcool}
The interaction is described by Lennard-Jones potentials
  \begin{equation}
  v_{\alpha \beta}(r) = 4\epsilon_{\alpha \beta}
  \left(\left(\frac{\sigma_{\alpha \beta}}{r}\right)^{12} -
  \left(\frac{\sigma_{\alpha \beta}}{r}\right)^{6}\right)
  \label{eq-LJ}
  \end{equation} 
The bond parameters $\sigma_{\alpha \beta}$ are $\sigma_{SS} = 1.05 $,
$\sigma_{SL} = 1.23 $, and $\sigma_{LL} = 1.21 $. These and all other
simulation parameters are given in reduced or Lennard-Jones units, and are
indicated by a star. The corresponding bond lengths are $2^{1/6}
\sigma_{\alpha\beta}$.
In overall outline, these distances are appropriate for a 
typical tetrahedrally-close-packed structure,
in which the small atoms have icosahedral coordination
shells and the large atoms have Frank-Kasper coordination
shells with larger  coordination number.
(Frank-Kasper structures are by definition 
``tetrahedrally close-packed (tcp)'', 
meaning that neighboring atoms always form tetrahedra
(Samson, 1968; Samson, 1969; Shoemaker et al.\ 1957; Frank and Kasper, 1958;
Frank and Kasper, 1959)\nocite {sam1,sam2,shoe57,FK1,FK2}.
This is not strictly true in our model -- or that of Henley and Elser
(1986) -- since there is a small number of non-tcp environments.)

The coupling constants were set to $\epsilon_{SS} = \epsilon_{LL} =
0.656$ for same species and to $\epsilon_{SL} = 1.312$ for different
species, in order to prevent phase separations into monatomic domains. 
To save computation time we cut off the potential 
smoothly at $r_{c} = 2.5 \sigma_{SL}$.
 
For the simulation we have modified the standard molecular dynamics method
to allow us to control pressure and temperature as described 
by Evans \& Morriss (1983). \nocite{eva}
A constant cooling rate may be introduced by the method described by
Lan\c{c}on \& Billard (1990). \nocite{lanbil}
The equations of motion are integrated in a fourth-order
Gear-predictor-corrector-algorithm (see for example Allen \& Tildesley
(1987)). 
The time increment $\delta t^{*}$ is
adjusted after testing for numerical stability. We
find that $\delta t^{*} = 0.00462$ is an appropriate value.
For simplicity the masses of small and large atoms are set to unity.

Our simulation box is a cube containing 128 small (S) and 40 large (L) 
atoms. The initial positions are set with a random number generator, 
and then relaxed. The generated liquid is equilibrated at an initial
temperature of  about $T^{*} = 1.0$. At high cooling rates we observe
a transition form a supercooled liquid to a glass at about $T^* = 0.5$. 
If the cooling rate is lowered to $T^{*}/t^{*} = 1.74\times 10^{-5}$ 
the supercooled liquid transforms sharply to an ordered solid at about $T^* =
0.6$, which analysis reveals is a quasicrystal (sometimes icosahedral and 
sometimes decagonal). The solid may still contain defects,
but the order can be improved by annealing. The annealed structures are
quenched down to $T^* = 0$ for structure analysis. Subsequent annealing runs
with $5 \times 10^{5}$ steps at constant temperatures of  $T^* = 0.5 $ and 0.4
show that 
the nucleated structures are stable in the sense that they do not change
except for defect annealing. More details about the cooling simulations can be
found by Roth et al.\ \citeyear{rotcool}.

\begin{figure}[ht!]
\centerline{\includegraphics[width=14cm]{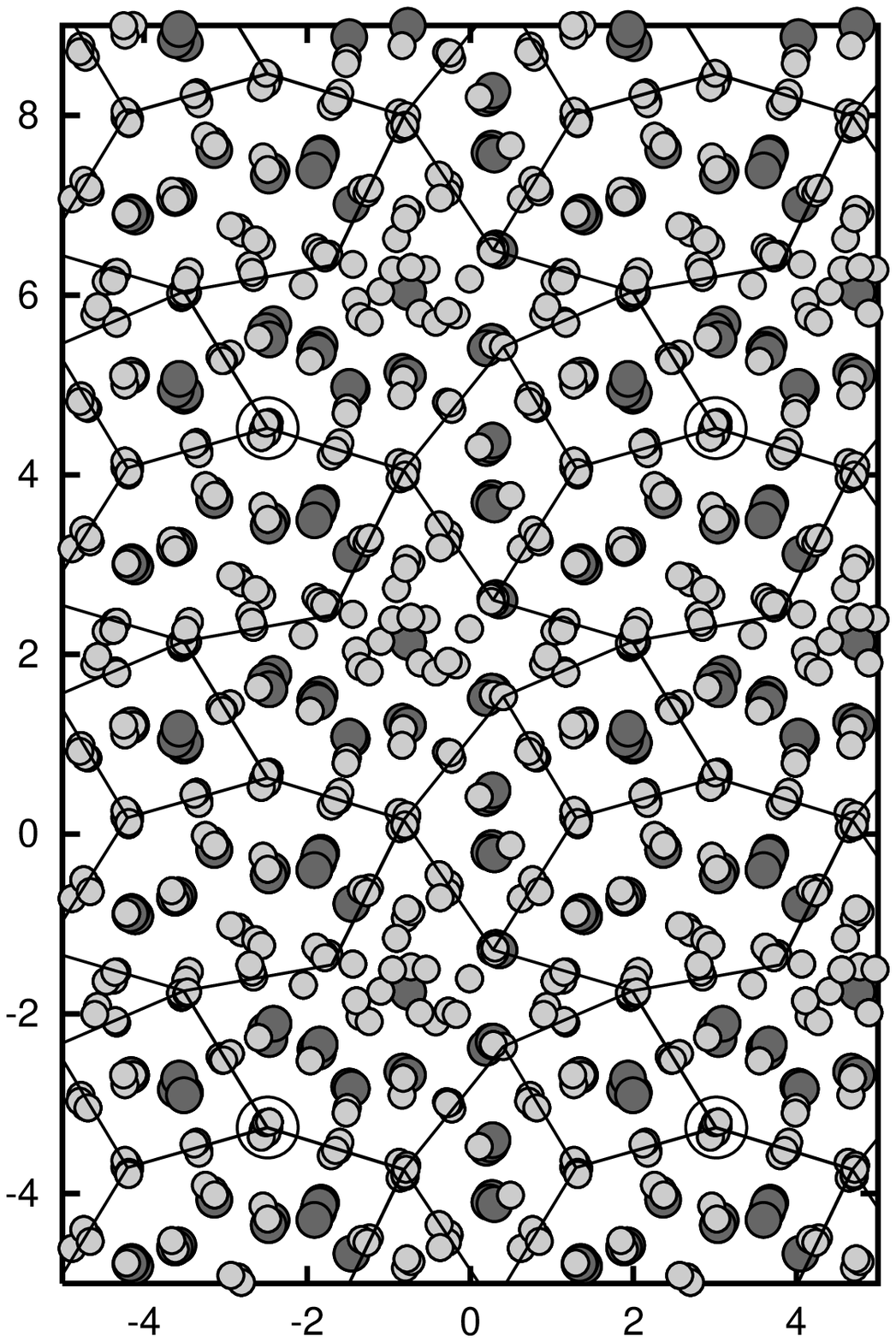} }
\caption{
Projection of the structure obtained by cooling down the `tenfold' axis. 
Lines link neighboring chains of interpenetrating icosahedra and outline the
tiles found in the structure. Four large circles outline the periodic boundary
conditions. The two types of atoms are indicated by shape and size.
}
\label{fig-proj}
\end{figure}

The new decagonal structure was observed at cooling rates of $T^{*}/t^{*} =
1.74\times 10^{-6}$ and $1.74\times 10^{-5}$.
The tenfold axis is aligned with the $\langle 1 1 0 \rangle$
diagonal of the cubic simulation cell; the structure is stacked periodically, 
with a stacking vector parallel to this direction.
Fig.~\ref{fig-proj} shows a projection of the simulation box onto a
plane normal to this axis; the atom positions are highly ordered and 
the approximant tenfold symmetry within this plane is obvious.
Because of the constraint of fitting into the fixed, periodic simulation cell, 
the structure is distorted  somewhat (in that the $x/y$ ratio within the
layers is squeezed relative to decagonal symmetry).

By examining slices transverse to the decagonal axis (not shown), 
we could distinguish three different types of
layers transverse to the tenfold axis, which we label ``A'', 
``B'', and ``X''. Let $c$ be the lattice constant in the stacking
direction; the simulation box extends over two periods ($2c$) before
encountering the same atomic layer. 
Most of the atoms in our structure are in A and B layers, 
which are separated by $\sim 0.8$ units.
The A layer (at height $0$) consists mostly of pentagons linked by
corners; the B layer (at $c/2$) also consists of pentagons, but these are
bigger and share edges instead. 
The basic motifs in the A/B layers are therefore pentagonal antiprisms.
Both the A and B layers contain a mix of L and S atoms. 
Between the A and B layers at $c/4$ and $3c/4$ are the much sparser X and $\rm
X'$ layers. 
These contain only S atoms centering the pentagons in the A/B layers, 
thus forming chains of interpenetrating icosahedra along the $c$ direction.
The X and $\rm X'$ layers are identical in projection on the
quasiperiodic plane.
The complete stacking sequence is $\rm AXBX'$.

In order to discuss the structure in greater detail, it is convenient 
to present an idealized model; we shall do this in the
next section.

The centers of the columns of icosahedra are shown linked by
added lines in Fig.~\ref{fig-proj}. 
Observe that the structures we found in simulations are not strictly
periodic in the stacking direction:
the axis centering a column of icosahedra may jump in position 
from one layer to the next. 

We also observed icosahedral structures
under the same simulation conditions.
Consequently, both phases might coexist in single (larger) sample. 
Such a coexistence is well-known
experimentally in the Al-transition metal structure family,
most commonly among metastable, rapidly-quenched quasicrystals  
but also for stable phase i(AlPdMn) and d(AlPdMn)
\cite{alpdmn}.
The simultaneous stability (or near stability)
of both structures suggests that the local order is very similar. 
Indeed, we shall show that our decagonal structure model is 
quite closely related to the well-known `Henley-Elser' 
\cite{henels} model.

\section { Decoration Models} 
\label{dsdm}

The most convenient framework for making sense of our structures, 
without any bias about the kind of long-range order, 
is a decoration of a tiling with atoms. 
The same decoration scheme, with a finite number of site types and
positional parameters, 
can model a random structure as a decoration
of a random tiling, or 
an ideal quasiperiodic structure as
a decoration of a quasiperiodic tiling.

However, in attempting to describe an idealized structure model, 
we suffer from an embarrassment of riches. 
That is, there are several related structure models describing
this kind of order.
The models differ by (i) breaking a local symmetry  in the position
of an atom (which introduces matching-rule-like  arrows) 
(ii) by allowing additional kinds of tiles
(iii) by constraining the packing of tiles (this
may create larger tiles as  composites of smaller ones).

Structure models are conveniently presented using idealized positions, 
represented as integer combinations of some finite set of basis vectors, 
but obviously the real atoms undergo displacements from these positions.
Two structural models are called
``topologically equivalent'' (Mihalcovi{\v c} et al.\ 1996a) \nocite{mihI}
when they converge to the same
structure after equilibration at low temperature 
(or at zero temperature by relaxation under the pair potentials. 
Although apparently  different, the models are physically
indistinguishable. 
Thus the concept of topological equivalence is crucial in 
comparing different structure models and in organizing the
above-mentioned family of related structure models.

We will begin (Sec. \ref{basictiles})
with a simple simple tiling of three tiles --
the fat and skinny Penrose rhombus plus distorted hexagon ``Q'' -- 
from which the other tilings can be derived. 
This is not itself a good tiling model, because the packing is
poor in the skinny rhombus, but it functions as a ``zero-order''
approximation for constructing better models.
Therefore, we go on to consider two different modifications
which correct this: 
(i) the triangle-rectangle tiling (Sec.~\ref{trirect}). 
and (ii) the ``two-level tiling'' (Sec. \ref{twolevel})
The relation of each modified model to the Penrose tiling decoration
is an example of a 
``differentiation'' in the terminology of Mihalkovi\v{c} et al.\
(1996b); \nocite{mihII}
this means that a single environment in one model corresponds to 
several subclasses in the second model, which have somewhat different
atomic decorations.

\subsection{ Basic Tiles}\label{basictiles}\label{basic}

Next we present the simplest possible version of the model,
phrased as a decoration of the Penrose rhombi without matching rules.

\subsubsection{ Penrose Rhombi}\label{penrose}

Consider the decoration of the fat and skinny Penrose rhombi, 
as shown in Fig.~\ref{fig-basic}(a,b).  
This structure is not the best packed, but it is certainly
one of the simplest.
Each large atom is at a position
dividing  the long diagonal of a fat rhombus in the ratio
$\tau^{-1}:\tau^{-2}$ where $\tau$ is the golden mean $(1+\sqrt 5)/2$. 
Notice that there is a complete symmetry
between  the A and B layers. Furthermore, the decorating
atoms sit on the midpoint of each tile edge, so the decoration
does not enforce the Penrose matching rules: it may be applied to
any random tiling of these rhombi. 

The centering atoms of the pentagons are called $\alpha$. They are located in
the X layer at the corners of the rhombi.
In the A layer,
the mid-edge atoms are called $\beta$
and the rhombus interior atoms $\lambda$.
In the B layer,
the equivalent mid-edge atoms are called $\gamma$
and the rhombus interior atoms are $\mu$.

There are two alternative criteria for assigning the species: either 
according to interatomic distances in the model, or according to 
coordination number \cite{henels};
the two criteria are obviously correlated and (for all
models presented in this paper) they lead to the same assignments.
We assign $\beta/\gamma$ mid-edge sites to be S atom, since
they have coordination 12 and tend to be squeezed by the
$\alpha$ atoms at either end of the edge.
The $\alpha$ sites are also assigned S atoms, since
they have coordination 12 and are squeezed along the $z$ axis 
where they form chains. 
(The interlayer spacing is less than an interatomic spacing, since
the other atoms don't form vertical chains.)
Finally, the $\lambda/\mu$ sites, which all gave coordination 16,
are taken to be L atoms. 

\subsubsection{ Q Tile }

In the Penrose decoration, described so far, 
the A and B layers are equivalent by a $10_5$ screw symmetry,
and correspondingly  $\beta$ is equivalent to $\gamma$ and
$\lambda$ is equivalent to $\mu$.
However, in the decorations we present later, the A and B layers will be
inequivalent.  In those more complicated structures, a single Greek
label refers to several subclasses of atoms having similar
but not identical local environments. (Usually their coordination
shells are topologically identical, but there is some variation in
which neighbors are L and S atoms.)

The Penrose tiling structure forces a ``short'' linkage between two vertices
when they are related by the short diagonal of the skinny rhombus. 
Here ``linkage'' denotes a connection between the centers of
neighboring motifs, which form the vertices of a rigid network.
(In this case, the motif is the chain of centered pentagons,
and the network is the tiling.)
The $\alpha$ atoms projecting to those two vertices
have {\it non}-icosahedral coordination shells. 
(There are similar to the awkward coordination shells occurring 
in the oblate rhombohedron \cite{henels}).
We believe these are unfavorable energetically.
(Indeed, nothing resembling a skinny rhombus is found in Fig.~1.)
Therefore, we shall make an additional hypothesis: 
{\it true chains of icosahedra are energetically favorable and 
so their frequency should be maximized}.

To increase the number of chains of icosahedra, we must somehow
eliminate the short linkages. 
This is possible whenever two skinny tiles and a fat tile
form  a distorted hexagon; they
can be joined in a composite tile Q 
as shown in Fig.~\ref{fig-basic}(c).
The only change in atomic positions in forming the Q tile is
along the axis over the central vertex, where the 
the $\alpha$ atoms forming the chain
are replaced by $\lambda'$ atoms 
(in A layers, if the adjacent atoms along the tile edges  are in a B layer,
and otherwise vice versa). 
Note there are half as many new
atoms $\lambda'$ as there were of old $\alpha$ atoms;
they should be L atoms because of this larger spacing 
(or alternatively, because they have coordination 15).
We will call this operation a ``chain-shift''.

Note that the Q tile is considered
to have a symmetry between the top apex and bottom apex
(the light dashed lines in Fig.~\ref{fig-basic}(c) 
only illustrate the topological equivalence to the
apparently asymmetric Penrose tile decoration.)
The Q decoration should have
the full symmetry of the Q tile's exterior;
after the chain-shift, this require only small adjustments of
position and species. 
The $\lambda$ site from the fat tile is renamed $\lambda'$
since it has coordination 15 after the chain-shift and is symmetry-equivalent
to the $\lambda'$ site from the internal vertex.
The non-Frank-Kasper $\gamma$ atom on the edge shared by skinny rhombi
is converted into a $\mu$ L atom (upper part of Fig.~\ref{fig-basic}(c))
just like the $\mu$ site from the fat rhombus.

Finally, the other two non-Frank-Kasper $\gamma$ sites  on internal edges
are renamed ``$\delta$''
and must be shifted slightly so as to lie
symmetrically on the bisector of the long Q diagonal
(since their neighbors are symmetric around that bisector.)
They divide the bisector line almost into thirds 
(we used $0.312:0.376:0.312$ in the initial  configuration
for simulations in Sec.~\ref{secopt}). 
Since the $\delta$ atoms have coordination 14, and are somewhat 
squeezed with their neighbors along the bisector line, 
they have been designated S atoms.
(However, note that
coordination $14$ was found to be
borderline in the closely related icosahedral structure:
analogous sites occurring on the short axis of the
rhombic dodecahedron  tile (see  below)
are occupied by small atoms in i(AlCuLi) but by large atoms in
i(AlMgZn)\cite{henels}).

To maximize the number of proper centered-pentagon chains, 
then, the original tiling should be arranged 
with skinny rhombi grouped in pairs wherever possible.
This is essentially the same mathematical problem 
as that of packing equal disks on Penrose tile vertices
\cite{hen86}; those the disks correspond to the 
pentagon-chain motif in the present model.

\subsubsection{ Icosahedral Tiling Relationship} \label{icorel}

\begin{figure}[ht!]
\centerline{(a)}
\centerline{\includegraphics[width=5cm]{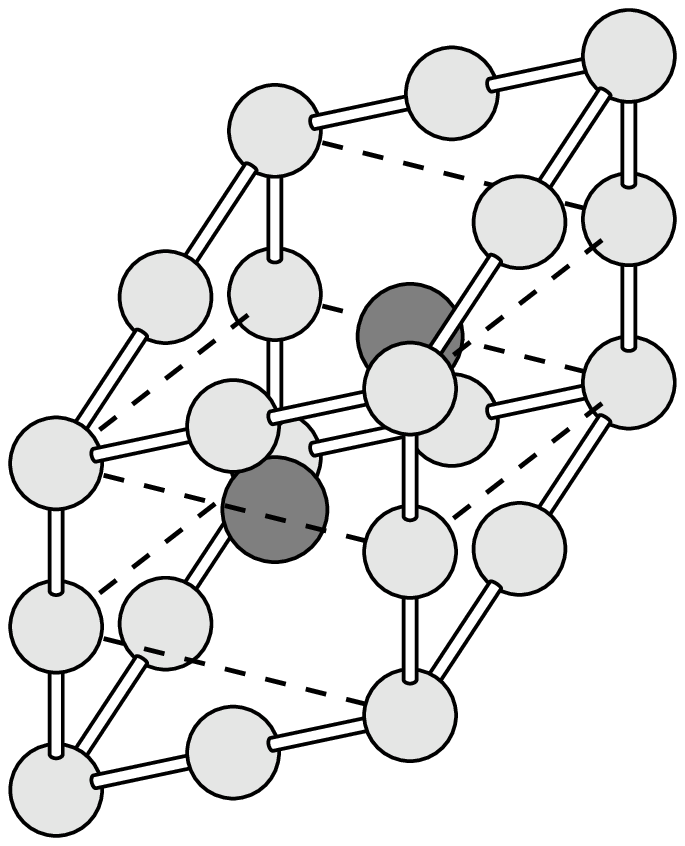} 
\includegraphics[width=5cm]{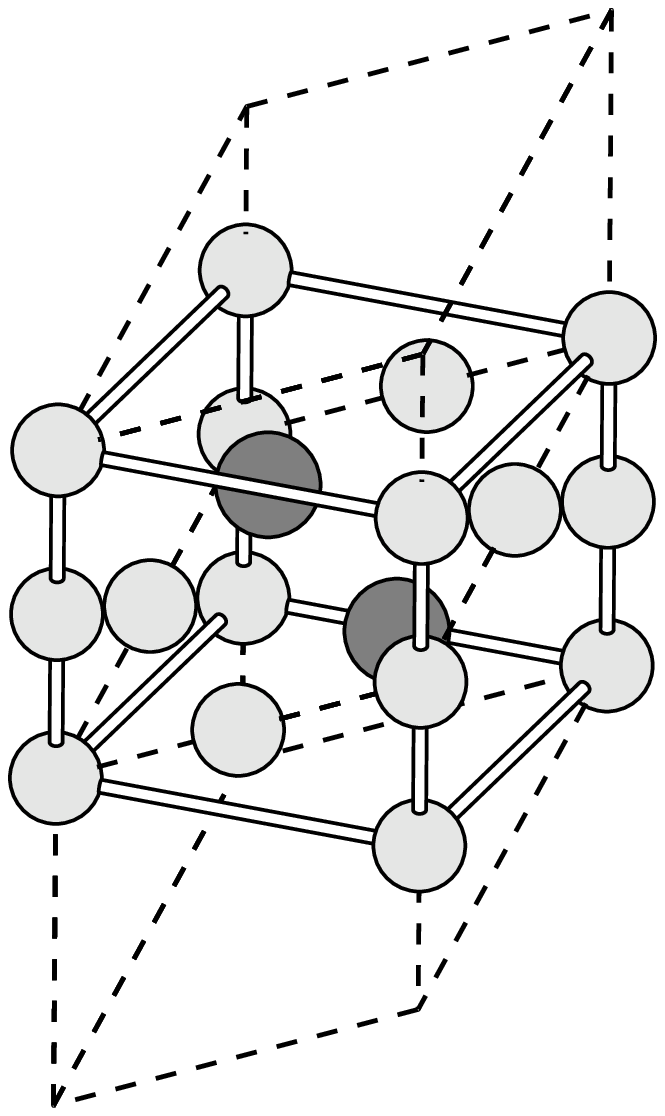}}
\centerline{(b)}
\centerline{\includegraphics[width=5cm]{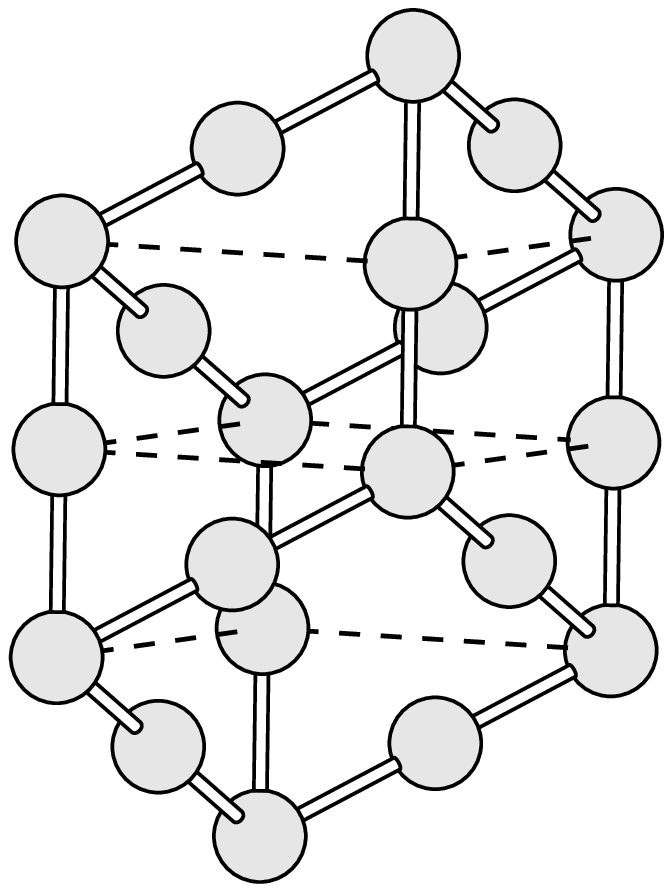} 
\includegraphics[width=5cm]{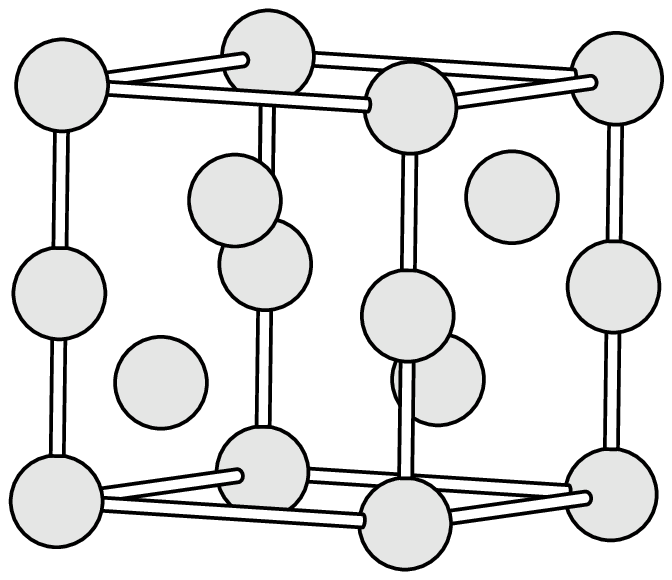}} 
\caption{
Rhombohedra with the Henley-Elser decoration;
dark spheres are L atoms and light spheres are S atoms.
Two cells are stacked above each other. The prisms are obtained by drawing new
bonds (indicated by dashed lines). 
(a) Prolate rhombohedron transformed fat rhombic prism; 
(b) Oblate rhombohedron transformed a skinny rhombic prism.}
\label{fig-PROR}
\label{fig-3D}
\end{figure}

It is interesting to note the close relationship of this tiling
to the three-dimensional ``Henley-Elser'' decoration of the 
Ammann rhombohedra by L and S atoms as a model of Frank-Kasper
icosahedral quasicrystals such as $i$(AlCuLi) and $i$(AlZnMg).
Let us orient rhombohedra so that one edge lies in the vertical direction
as shown in Fig.~\ref{fig-PROR}, but shear them slightly so that
the vertical edge has length $c$ and the other five kinds of 
edge have vertical component $c/2$ and horizontal components
of length $a$ along one of the five basis vectors. 

Projecting the prolate and oblate rhombohedra onto the plane 
perpendicular to the selected axis generates respectively the 
fat and skinny rhombus of the (2D) Penrose tiling.
Thus, every tiling of the rhombi can be extended 
vertically as a stacking of rhombohedra with period one edge length
along the special axis.
The layer of this stacking is a puckered slab dual to a grid-plane,
analogous to one of the ``tracks''
in a two-dimensional Penrose tiling \cite{dunkat}. 
Note that, in contrast to a generic rhombohedral tiling, 
{\it every} rhombohedron in this stacking has edges 
in the vertical direction.

In fact, it turns out that our decagonal decoration  of Penrose tiles
is ``topologically equivalent'' to the icosahedral decoration
of rhombohedra by Henley \& Elser (1986) \nocite{henels} 
\footnote {A similar, but quite imperfect,  
relationship can be made between the
icosahedral $i$(AlMn) (Elser \& Henley 1985; Mihalcovi{\v c} 1996b)
\nocite{elshen,mihII} 
and the decagonal  model based on the Al$_{13}$Fe$_4$ structure
\cite{hendec}}.

To see this, cut the rhombohedra along horizontal planes such 
as those shown dashed in Fig. 2(a,b) 
-- all the atoms in the PR and OR lie in or near these 
rather densely packed layers.
Then reassemble them into prisms with horizontal rhombic faces.
The atom positions form a decoration of the prisms, 
shown in Fig.~\ref{fig-3D}(c,d), 
which is clearly equivalent to that in Fig.~\ref{fig-basic}(a,b).
(The ideal positions by Henley \& Elser (1986), \nocite{henels}
projected onto the plane, 
are {\it exactly} those used  here in the tiling; 
the vertical positions of the $\lambda$ atoms, however, 
must be adjusted by $\sim 0.026 c$.)

\begin{figure}[ht!]
\centerline {(a)}
\centerline{\includegraphics[width=10cm]{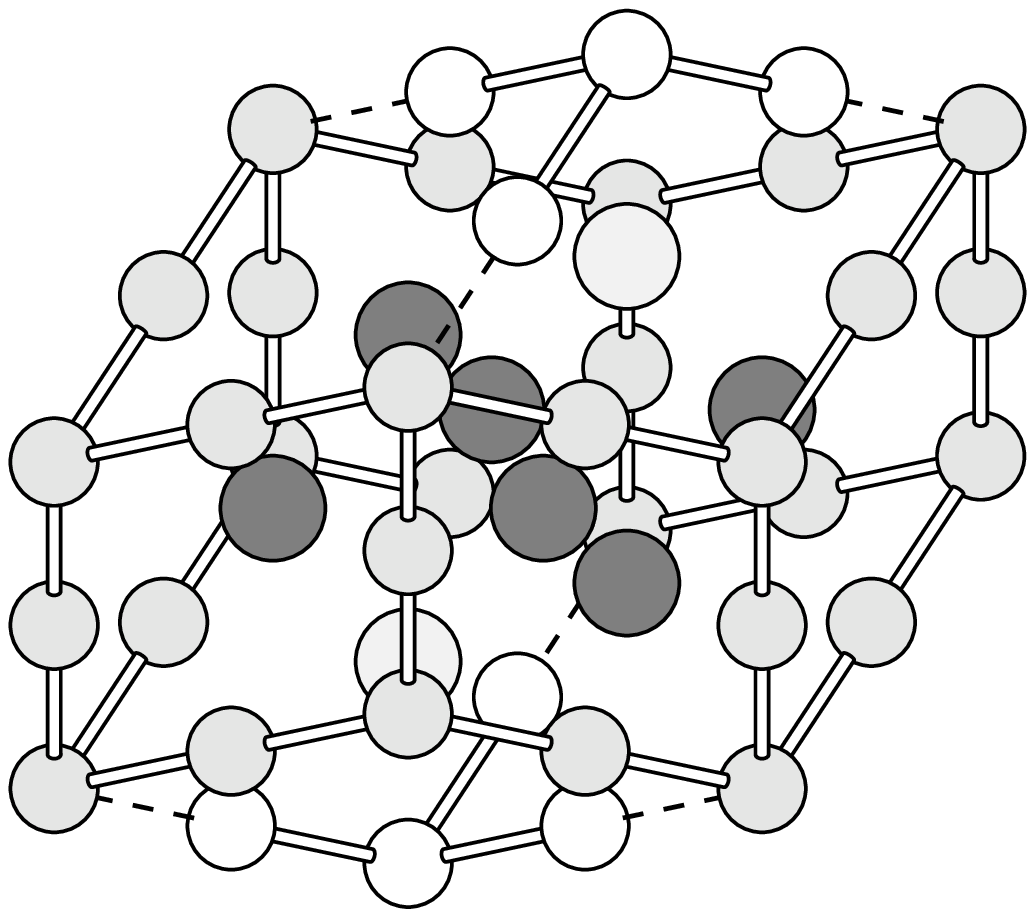}}
\centerline{(b)}
\centerline{\includegraphics[width=6cm]{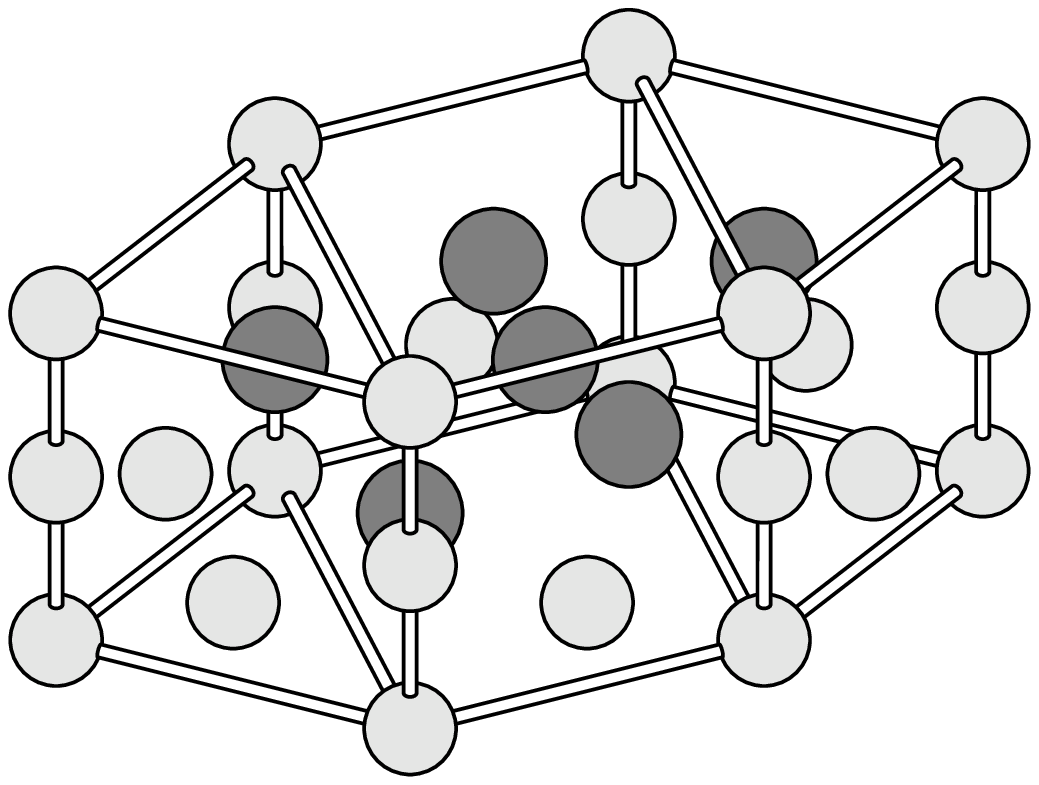}}
\caption{
Transformation of rhombic dodecahedron (RD) into 
decagonal Q tile. 
(a) The white spheres have to be removed.
(b) After dividing into layers, the modified RD is reassembled
into a hexagonal prism.}
\label{fig-RD}
\end{figure}

\begin{figure}[ht!]
(a)\includegraphics[width=6.5cm]{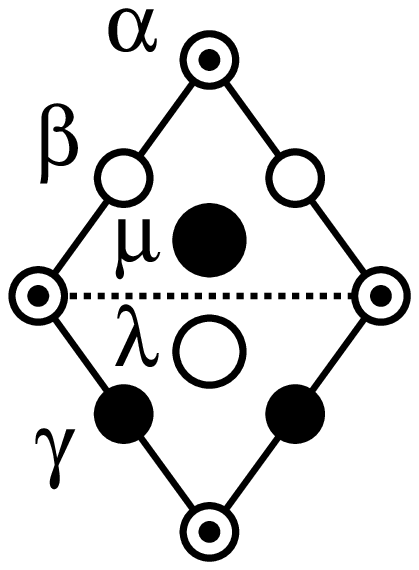}
(b)\includegraphics[width=6.5cm]{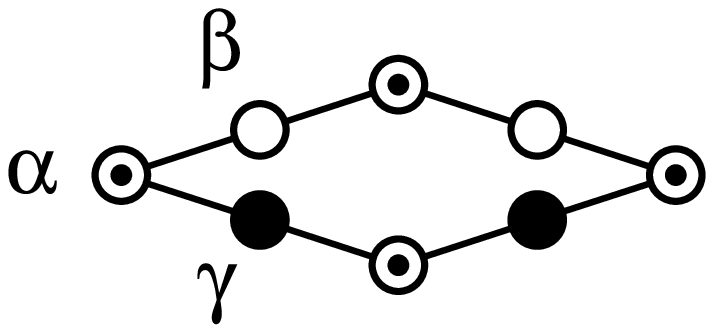} 
\centerline{(c)\includegraphics[width=5cm]{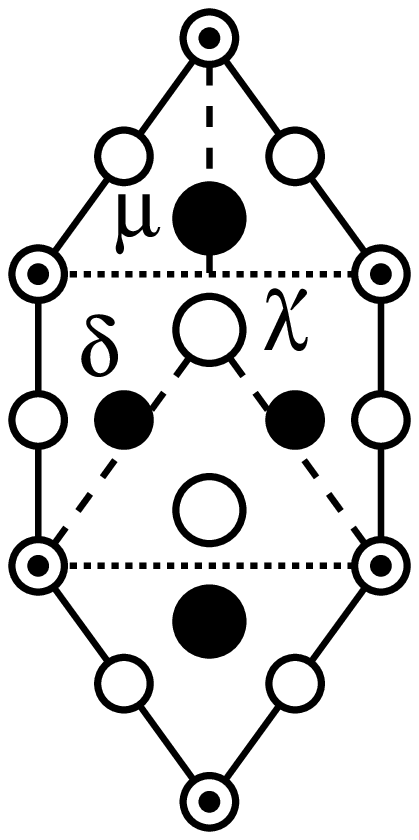} \hfill
(d)\includegraphics[width=5cm]{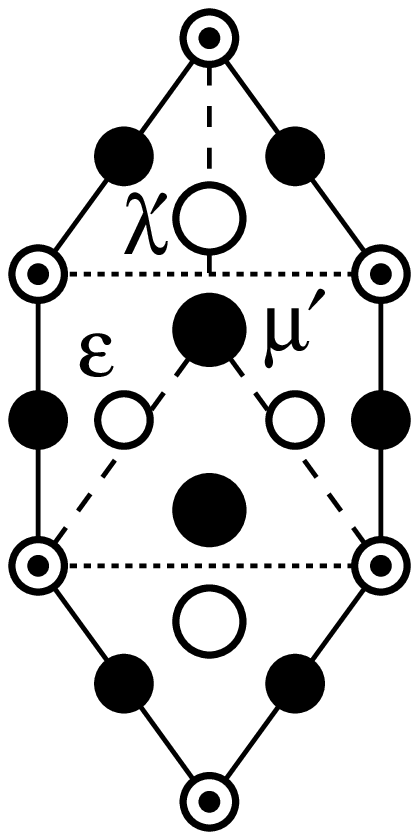} }

\caption{ Basic tiles and basic decoration.
The figures show a projection, with atoms marked by circles; 
atoms in A,B, and X layers are white,black, and centered
circles respectively (see key); large (L) atoms are  shown larger.
Site classes are labeled by Greek letters. Dotted lines indicate the
decomposition of the fat rhombus and Q tile into triangles and rectangles.
(a) Fat Penrose tile
(b) Skinny Penrose tile
(c) Q tile.
(d) Q tile in the fR/Q tiling.
Light dashed lines divide it into three Penrose rhombi (used only for
construction);  
dashed double lines show $b$ edges, dividing it
into a rectangle and two isosceles triangles.
The ``$\delta$'' atoms would be relaxed outwards in a more realistic model.}
\label {fig-basic}
\end{figure}

The third tile of the Henley-Elser model, the
rhombic dodecahedron (RD), is a little bit more complicated. 
(see Fig.~\ref{fig-RD}). Before it can be stacked, 
two skew triangular prisms at opposite sides of the RD, forming together one
PR, must be removed. 
If this truncated RD is cut into layers and reassembled, 
it produces an elongated hexagonal prism which 
is decorated exactly as the Q tile in Fig.~\ref{fig-basic}(c). 

The structure obtained by transforming the Henley-Elser structure
not only has the atoms in exactly the same positions (after
the small vertical shifts) as our decagonal model, 
but also has exactly the same assignment of S and L atoms.
That is not very surprising, since the species were assigned
according to the local environments in both models.

\subsection{ Triangle-Rectangle Tiling}
\label{trirect}

Returning to the 2D tiling, 
there are now two possible schemes to handling the 
skinny rhombi everywhere in the tiling. 
In one scheme, we imagine first that {\it all} skinny tiles are 
paired and absorbed into Q tiles: a generic tiling of this sort is
the ``fR/Q (fat rhombus and Q) tiling''.

Then the Q tile can be divided into
a rectangle (R) and two isosceles triangles (T), with new
edges of length $b\equiv (1+\tau^{-2})^{1/2} a$, 
while each rhombus
can be divided into two isosceles triangles of the same kind. 
The two objects can form random tilings of the plane,
most of which cannot be regrouped into rhombi \cite{ox}.
The decoration introduced in this paper is compatible with an 
arbitrary triangle-rectangle (T/R) tiling.
Note that, in this variant of the structure model,
each tile comes in an A and a B flavor,
denoting the level at which its mid-edge atoms sit;
tiles adjoining each other by a $b$ edge have opposite flavors. 

The triangle and rectangle have long been known as the building blocks for the
so-called pentagonal Frank-Kasper phases (see for example
Samson (1968); Samson (1969)). \nocite{sam1,sam2}
The 12-fold symmetric quasicrystals, which are typical Frank-Kasper phases,
are built from a triangle and square
on which the atomic arrangements are similar, but not identical, 
to those in the T and R tiles in our model.
Furthermore, a T/R description like ours was used by 
He et al.\ \citeyear{chinese} 
to describe
the $\rm Fe_{52}Nb_{48}$ quasicrystal, the only real 
material claimed to be a Frank-Kasper decagonal.
(See Sec.~\ref{comparison};
Nissen \& Beeli (1993)
\nocite{beeli}
also used a rhombus-triangle-rectangle tiling
to analyze TEM images of a Fe-Nb decagonal quasicrystal.)
However, the decagonal T/R structure has been systematically
studied as an abstract (random) tiling only recently
\cite {ox,cock}.

We shall now consider energy energy differences between different
T/R tilings which might cause additional ordering.
It would seem unfavorable energetically for two rectangles to 
adjoin along an $a$ edge, 
since the $\delta$ and $\beta/\gamma$ atoms along the bisecting axis
are tight within each tile and would be squeezed against each other;
let us add this to the constraints defining our tiling ensemble.
Then Cockayne's quasiperiodic T/R packing\cite{cock}
satisfies the constraint just mentioned; it satisfied a
second constraint, in that rectangles 
don't adjoin along $b$ edges, either.
With the latter constraint, every rectangle must
be capped with triangles forming a complete Q tile, and then all other 
triangles must be paired along their $b$ edges forming complete fat rhombi:
thus the structure is more economically described as an
fR/Q tiling with the constraint that Q's may not lie side-to-side.
(We presume the ensemble of such packings is a random tiling, but this
is not proven.)
Every fR/Q structure, of course, can be considered as a fat and skinny (fR/sR)
rhombus tiling produced by decomposing the Q tile. 
It has been pointed out \cite{mihox,ox}
that an fR/Q tiling gives the
optimal disk packing 
(i.e. maximal density of $\alpha$-chains, in our models)
of all Penrose tiling, since
{\it all} the skinny rhombi are grouped into pairs, unlike the two-level tiling
described below.

Possibly the additional constraint on the T/R tiling 
is justified by the energetic cost of adjoining along the $b$ edge --
the $\lambda'$ atoms inside one R press against
the $\delta$ atoms (in the same layer) inside the other R --
but we have not identified any compelling argument and hence
do not propose the fR/Q tiling as the optimal geometry.

\subsection{Two-Level Tiling} \label{twolevel}

\begin{figure}[ht!]
\centerline{\includegraphics[width=10cm]{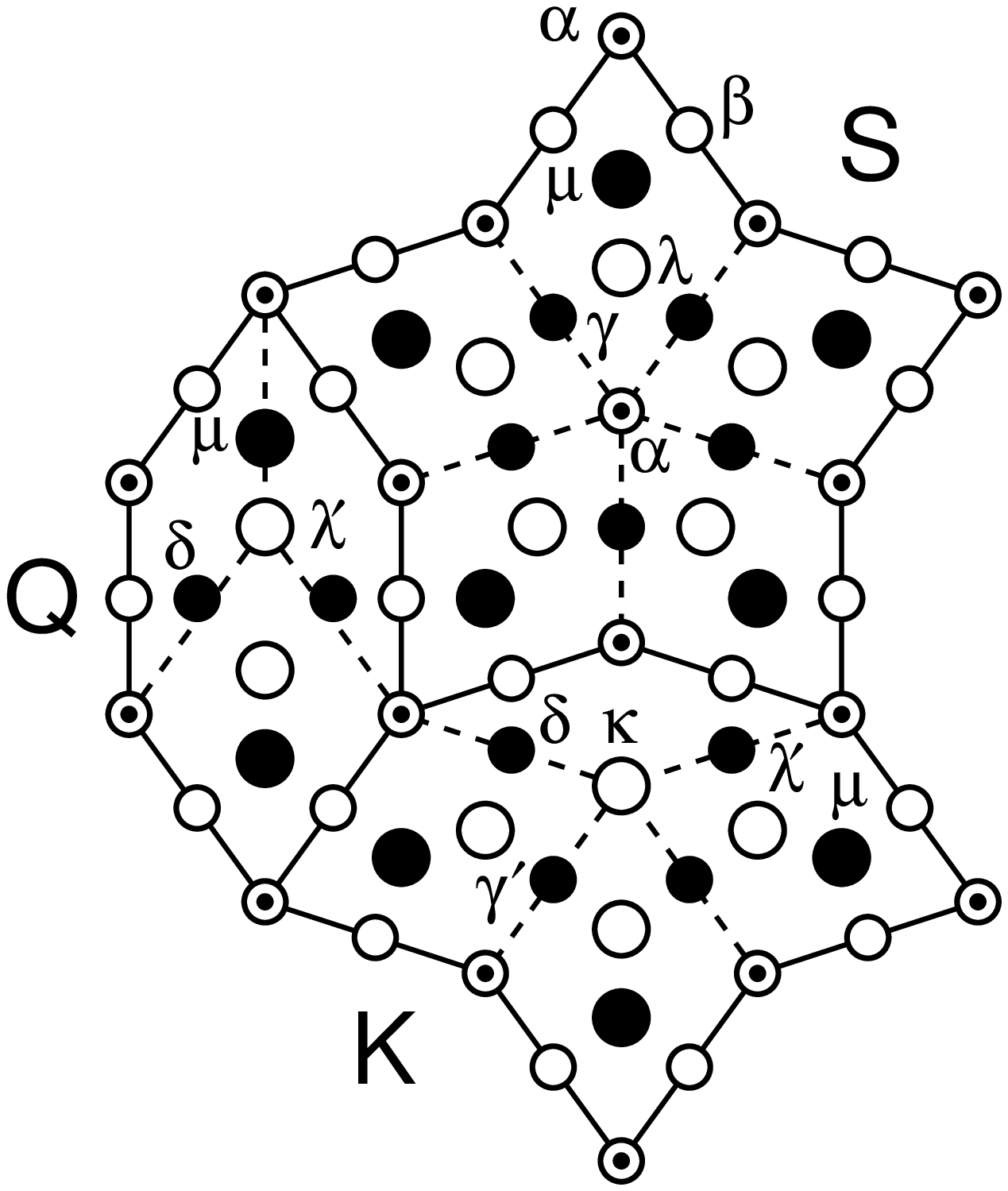}}
\caption{ Decoration of the two-level tiling. 
Tile edges are shown solid; dashed edges show how they
are combined from Penrose tiles.
}
\label {fig-2LTdec}
\label {qksdecl}
\end{figure}

The decagonal structure observed after cooling from the 
melt (Sec.~I) had the additional property
that its A layer is inequivalent to its B layer.
We shall now turn to a structure model which adopts this
as a postulate.
Chain-shifts occur only at vertices where all the edges are decorated with 
B layer atoms. Naturally, the B atoms in turn shift away from the midpoint
of the edge towards the partially vacated columns. (This was observed
in the simulation, in the fact that the B layer pentagons were bigger
than A layer pentagons.)
This has the effect of the double-arrow matching rule in the Penrose tiling. 
An arbitrary tiling obeying only the double-arrow matching rules 
(Tang and Jari\'c, 1990) is equivalent to a random tiling
of three composite supertiles Q, K, and S\footnote{
The three tiles are called ``Q, K, S'' after the corresponding 
three local environments found in the Penrose tiling where double-arrow
edges can meet. The same tiles are called ``h,c,s'' respectively 
by Li \& Kuo (1991).
\nocite{li2}
}
which are obtained by merging Penrose tiles along the double-arrow edges.
Since the atoms decorating the outer edges of the Q, K, and S tiles
remain symmetric about the middle of the edge, 
they do {\it not} enforce the (single-arrow) Penrose matching rules 
on these edges: all random tilings of the plane by Q, K, S tiles
are a priori permissible.
The random tiling of these tiles has been called the 
``two-level'' tiling (2LT)\cite{henART}; these tiles were independently
proposed by Li \& Kuo (1991). \nocite{li2}
Fig.~\ref{fig-2LTdec} shows a fragment of the 2LT, including one of each tile.
As a special case, the Penrose tiling,  
modified by combining rhombi along their double-arrow edges, 
becomes a simple quasiperiodic Q, K, S tiling. 

The ``Q'' tile decoration was described already; 
the star tile S is also straightforward, being a composite of five fat rhombi.
Note that 
no chain-shift is necessary at the center of the S tile 
since there is no skinny tile adjoining; 
thus this site forms an exception, around which the A/B symmetry breaking 
(between large and small pentagons) is opposite to the pattern around all other
$\alpha$-chains.
Otherwise, all the edge sharing pentagons are in layer A and the 
corner sharing pentagons are in layer B. 

The ``K'' tile is a grouping of a skinny rhombus
with three fat rhombi; there is a chain-shift on the central vertex, which
becomes known as a $\kappa$ site.
This is the second option for absorbing the extra 
single skinny rhombi, instead of pairing all of them in Q tiles.
However, the atomic arrangements in the K are still somewhat awkward. 
The four $\gamma$ atoms surrounding the internal vertex
can only relax to form a sort of pentagon with a missing corner,
meaning that the $\kappa$ coordination shell 
is not a Frank-Kasper polyhedron but a distorted octahedron.

We could have arrived directly at the two-level tiling
from analysis of Fig.~\ref{fig-proj}, if we
accepted as a postulate that in the linkage between 
two chain-motifs, the edge-sharing pentagons are always in the B layer.
We quickly arrive at a geometry where the 
angles between linkages are all multiples of $2\pi/5$. 
The smallest tiles enclosed by such edges are the Q, K and S 
tiles of the ``two-level'' tiling (Fig. \ref{fig-2LTdec}).
The details of this derivation are in Appendix A.

Reviewing Sec.~\ref{dsdm}, 
we can identify a hierarchy of successively more ordered tiling geometries.
The Penrose tiling is the most ordered;
it can be broken up into 
the ``two-level'' tiling of Q, K, S tiles. 
The K and S tiles can be broken up further into fat and skinny rhombi; 
finally the fat rhombi  can be broken into triangles, and the
Q tiles broken into triangles and rectangles.

\section{ Quasiperiodic Models and Stoichiometry}
\label{environments}

Even when a quasicrystal model has been formulated as a tile decoration,
it is often illuminating to represent it
as a quasiperiodic structure obtained by a cut 
through a higher-dimensional space, 
since
(i) the space-group and diffraction properties are most concretely
understood in this fashion
(ii) different site types are represented as domains of the 
acceptance domains in the ``perp'' space, thus
the relations between different site types are
visible in the spatial relation of the domains.
(iii) The number density (or frequency) of a particular species or site type
is proportional to the area of its portion of the acceptance domain,
which is convenient for computing a definite stoichiometry.

This description is helpful even for atomic models based on
random tilings. These, when ensemble-averaged, produce structures of 
partially-occupied sites. Mapped into perp space, such
distributions can usually be {\it approximated} as 
quasiperiodic acceptance domains convolved with a Gaussian 
in perp space\cite{henART}.

The first step is to find a quasiperiodic tiling.
This is trivial for the two-level (Q/K/S) tiling, but highly nontrivial
for the T/R (or for the fR/Q) tiling; in that
case, the tiling can only be constructed by an elaborate deflation, 
and the corresponding acceptance domains have fractal boundaries
\cite{cock}.
Therefore, we will limit ourselves
to presenting the acceptance domains based on the Q/K/S-tiling decoration,
where the tiles are
arranged in the perfect Penrose tiling. 
(A Penrose tiling is turned into a two-level (Q/K/S) tiling simply by
erasing the double-arrow edges).

Also in this section, we tabulate the contents in atoms of each tile.
This approach to stoichiometry is a more powerful than the approach using
the acceptance domain, since it can be applied even to random tilings.\\[1ex]

\subsubsection{ Idealized Structure and Acceptance Domain}

In order to construct the acceptance domain, 
we first make a crude idealization
of the Penrose-rhombus decoration which we presented first
(Fig.~\ref{fig-basic}(a,b)).
All $\beta$  atoms lie on single-arrow edges
and all $\gamma$ atoms lie on double-arrow edges.
Let us shift the $\beta$ atoms and the
$\gamma$ atoms in the direction of Penrose's arrows, 
such that each divides its edge
in the ratio $\tau^{-1}:\tau^{-2}$.
(This ratio is technically convenient in allowing
the ideal coordinates to be written as integer combinations
of the five basis vectors.
They are actually most simply treated using a
deflated Penrose tiling basis with lattice constant $\tau^{-1} a$.) 

In going to the Q/K/S tiles for in the quasiperiodic structure,
we retain these same positions but change the label and
the species as described in Sec.~\ref{twolevel}; note in particular
that the displacement properly places the $\delta$ atoms on the
bisecting symmetry axis of the Q tile.
The B layer forms Penrose's packing of regular pentagons, one 
centered on every vertex of the 2LT.
Each ``internal'' vertex (where double-arrows meet) is also surrounded
by a pentagon, but these are irregular and have many short distances.

The acceptance domains are shown in Fig.~\ref{fig-acc}.
Note there is a separate domain for layers A, B, and X of the real structure;
in addition, as always, the vertices project into five flat two-dimensional
layers in three-dimensional perp-space, so we must show separate
shapes for each perp-space layer.
Finally we have shaded the domain with dark and light shading to distinguish 
L and S atom occupancy.

Although this structure  has unreasonable distances, it is 
``topologically equivalent'' (Mihalcovi{\v c} et al.\ 1996a)\nocite{mihI} to
the 
(reasonable) version presented in figure \ref{qksdecl}.
We could alternatively compute the acceptance domains for the
decoration in figure \ref{qksdecl}, 
with the atoms moved to mid-edge positions (except for those in the Q 
tile). 
Relative to Fig.~\ref{fig-acc}, 
the  $\beta$, $\gamma$, and $\delta$ subdomains 
each would get broken up into five pieces, 
and shifted in the parallel space direction. 
Since the mid-edge atom positions are also
special crystallographic positions, several of the shifted subdomains
will reunite there and form a new piece of acceptance domain
which is centered on a mid-edge site in the 5-dimensional hypercubic lattice.
We forgo to present these domains since this second
picture is less compact and not as appealing as the first one.

The space group of this quasiperiodic decoration is $P10 /m m m$. In other
variant decorations in which the 
A/B symmetry is not broken, we would 
get an additional system of mirror planes perpendicular to
the 10-fold axis and this changes the space group to $ P10_{5} /m m c$. 
The point
group of the motif with the highest symmetry is $5 /m m$
\cite{hendec}. 

\subsubsection{ Stoichiometry}

The most general approach to stoichiometry is to count the number of atoms
associated with each tile. 
Table I summarizes the contents of each tile.
The combination Q+S has contents 29L+14S  while 2K (which covers
the same area as Q+S) has contents 28L+14S.
This suggests that the K tile is too loosely packed;
indeed, the average coordination number in K is also a bit low.

Table II gives the coordination numbers and the net stoichiometry for each
decoration. 
In the case of the 2LT (Q/K/S) decoration, this agrees with 
the stoichiometry which can be read off from Fig.~\ref{fig-acc}.
It can be seen that the stoichiometry is close to $L S_2$ in both cases.
(The pure Penrose tiling decoration, an unrealistic model, 
has $L_{0.198}  S_{0.802}$.)
That seems to be considerably fewer L atoms than in the T(AlZnMg)
icosahedral approximant \cite{berg}
or Samson's AlMg-type structures, but it is comparable
to i(Al$_6$CuMg$_3$) in which 
only $30\%$ of the atoms are L (Mg) \cite{sam1,sam2}.

\section{ Further Simulations}
\label{XXX}

In section III, we worked out the geometric properties of the idealized model, 
in several closely related variants based on different tiling rules.
We derived the coordination configuration for all sites:
all atoms have a reasonable distance from their
neighbours and most of them are tetrahedrally close packed. 

The idealized model, in any of the variants presented in 
the last two sections, is well-packed 
and (nearly) tetrahedrally close-packed around every atom.
Hence it is expected to be (locally) stable against disordering
by thermal fluctuations.
To check this stability, we ran additional simulations
starting with a configuration from the ideal model. 

A major motivation of these simulations was to locate the 
region of parameter space in which the quasicrystal phase may be
thermodynamically stable; hence, we attempted to adapt the 
potentials to the structure model, as explained in Subsec.~\ref{secopt}.
As a result of the optimization there 
we can predict the ideal c/a ratio
and the density of the structure, as well as
the ratios of bond radii for LL, LS and SS pairs  which seem
most conducive for (decagonal) quasicrystal-forming.

We used not only Lennard-Jones (LJ) potentials as in eq.~(\ref{eq-LJ})
but also the Dzugutov potential, with independent parameters for 
the three ways of pairing species $\alpha$ and $\beta$.
Both pair potentials have a single attractive well with a minimum at radius 
$2^{1/6} \sigma_{\alpha\beta}$, 
and having depth $\epsilon_{\alpha\beta}$; 
here $\sigma_{\alpha\beta}$ and $\epsilon_{\alpha\beta}$
are called the ``bond parameter'' and ``interaction parameter''.
The Dzugutov potential \cite{dzug}
has an additional repulsive maximum at a radius $\sim 1.6$
times that of the minimum, 
and a height $\sim 0.5 \epsilon _{\alpha\beta}$;
this is designed to disfavor the square arrangements found in 
fcc, bcc or hcp structures.

\subsection{ Optimization of Parameters}
\label{secopt} 

To find the optimal potential parameters, we minimized the potential energy
at $T=0$ while varying these parameters. For this purpose we used a fixed
ideal structure model with the ideal atom position and without relaxation.
The sample size was a $p/q~=~3/2$ approximant\footnote{ 
the basis vectors are $(2 q, 0), (p-q, p), (-p, q), (-p, -q), (p-q, -p)$, and
the periods are\\ $2(2 p-q) \tau+2(3 q-p)$ in $x$-direction and $4
\sin(\pi/10)(p \tau+q)$ in $y$-direction.} of a perfect Penrose tiling
with 10 periods of XAXB layers and ideal size $l_x \cdot l_y \cdot l_z = 18.95
\cdot 16.09 \cdot 22.93$, containing 3270 small and 1420 large atoms. Samples
of different size do not yield different results within the accuracy we
achieve. 

In a three-dimensional structure with two species of atoms L and S, 
bond and interaction parameters  
$\sigma_{\alpha\beta}$ and $\epsilon_{\alpha\beta}$ must each be determined
for  pairs $\alpha\beta$ equal to $LL$, $LS$, and $SS$.
Of course, one of the $\epsilon_{\alpha\beta}$'s 
can be eliminated in principle, 
in the choice of the energy  unit.
We chose the sample size/shape (which was kept fixed in the $xy$ plane
of the layers) so as to 
the Penrose tile edge length at exactly $a=2$
(the nearest-neighbour distance is roughly $a/2$.)
Thus all three $\sigma_{\alpha\beta}$'s are nontrivial parameters.

Results obtained in optimizing potentials for binary icosahedral Frank-Kasper
structures (Roth et al.\ 1990)\nocite{rotstab} with similar local structure
and potentials, show that 
$\sigma_{\alpha\beta}$ are largely independent of
$\epsilon_{\alpha\beta}$, for $\epsilon_{\alpha\beta}$ in the range of $0.5 <
\epsilon_{\alpha\beta} < 2.0$. 
Therefore we set $\epsilon_{LS} \equiv  1$ and $\epsilon_{LL} = \epsilon_{LS}
\equiv 1/2$. Thus, we actually varied only the three
$\sigma_{\alpha\beta}$'s. 

Before we can carry out the optimization itself we have to determine the
lattice constant $c$ in the stacking direction. 
(The A/B layer spacing $c/2$ should also be roughly a lattice constant.)
This is done in a poor man's minimization by
scanning the interval $0.8 < c < 4.0$ and looking for the minimum of the
potential energy $E(c)$ in the following way: 
having fixed $a$ (and consequently $l_x, l_y$) 
we calculate the
partial radial distribution functions $g_{\alpha\beta}(r)$ 
for the ideal model using a given layer distance $c$. 
Then, without annealing or relaxing, we compute the partial potential energy
  \begin{equation}
   E_{\alpha\beta}(\sigma_{\alpha\beta}, c) = \int_{0}^{r_{c}} v_{\alpha\beta}
   \left(\frac{r}{\sigma_{\alpha\beta}}\right)\, g_{\alpha\beta}(r, c)\,  4\pi
     r^{2}\, dr 
  \end{equation}
dependent on the cut-off radius $r_{c}$ and the pair potential
$v_{\alpha\beta}$ and minimize it as a function of
$\sigma_{\alpha\beta}$.
Using this procedure we find that the total potential
energy $E = E_{LL} + E_{LS} + E_{SS}$ is minimzed if each
$E_{\alpha\beta}$ is optimized separately.

The optimal $\sigma_{\alpha\beta}$ for Lennard-Jones potentials are
$\sigma_{SS} = 1.029$, $\sigma_{SL} = 1.137$, and $\sigma_{LL} = 1.189$.
For the potentials used by Dzugutov we get $\sigma_{SS} = 1.108$,
$\sigma_{SL} = 1.034$, and $\sigma_{LL} = 0.929$.
The result reflects the fact that the Dzugutov potentials are very
short ranged and that they have a maximum repulsive for second nearest
neighbours, whereas even in the cut-off and smoothed version of the
Lennard-Jones potential interactions with the second or 
third set of neighbours are still attractive.

We also found that the optimal A/B interlayer distance for Lennard-Jones
and for Dzugutov potentials is about $c/2= 1.04\pm 0.01$
The uncertainty quoted here 
reflects the slight variation of the spacing depending on
the choice of  $\epsilon_{\alpha\beta}$ and $r_{c}$.

\subsection{ Results}
 
The constant temperature and pressure molecular dynamics
simulation method described in Sec.~\ref{simulations} has been applied to
study the 
thermodynamic stability of the ideal structure. To simulate non-cubic
structures 
it has been extended to allow independent changes of the box lengths $l_{k}$.
Thus during the MD simulations the box size could vary in contrast to the
optimization, were the box was fixed.
The simulation sample was the same as the one used in Sec.~\ref{secopt}.
With Dzugutov's potentials the melting temperature at $P^{*} = 0.01 $ is
$T^{*}_{\rm m} = 1.23$. (In fact, the structure does not melt; it
vaporizes.)

Of course, in an infinite box the ratio $l_x/l_y = 2 \sin 36^\circ \sim 1.1756$
is fixed by pentagonal symmetry. In a finite system, however, 
$l_x/l_y$ should deviate
slightly from the ideal ratio due to the ``phonon-phason'' coupling
(Lubensky, 1988, and references therein) 
since the periodic boundary conditions force a nonzero 
phason strain.
We found the equilibrium value $l_y/l_x = 1.177$ in the simulation, 
independent of the temperature and very close to the ideal value.

The A and B layers are mirror planes so they should still be flat in the
relaxed structure; on the other hand, the X layers should pucker slightly
after relaxation, with the $\rm X'$ layer puckered in the opposite direction.
The simulation, however, show that the layers remain essentially flat.

We have also recorded the atomic mobility by monitoring the mean square
displacement of the atoms. It turns out that even close to the melting point
($T = 0.9 T^{*}_{\rm m}$) 
long-range diffusion is {\em not} observable at all,
and only very few atoms are seen to jump to a new position.
Those atoms which do jump are all in the skinny rhombus part of the K-tile.
Similar results have been obtained for icosahedral and dodecagonal
structures \cite{rothdiff} where oblate rhombohedra and skinny
rhombic prisms resp.\ play the same role as the flat rhombus in the K-tile.

Most of the jumps in the K tile take place around the $\alpha$ atom at the
top of the tile (see Fig. \ref{qksdecl}) and around the interior vertex.
Possibly the jumping in this environment indicates that our model is erroneous 
there:
perhaps a different arrangement or choice species should occur
in the atoms surrounding the interior vertex of the K tile, or perhaps
the correct model is a T/R tiling which has no K tiles
at all.
It is also possible that the correct structure model should be thought
of simply as a random tiling of Penrose rhombi
(Sec. \ref{penrose}); 
then ``phason fluctuations'' are realized by tile flips
that entail switching which vertex of a skinny rhombus undergoes a
``chain-shift''. 

Our conclusion is that the structure is very stable, and that the
choice of the atoms positions is reasonable. The jumping atoms do not affect
the stability since they are separated by large distances.

\section{ Discussion}

\subsection{ Summary}

Now we are back at the beginning: we started with a structure found by
computer simulations, and recognized that the structure could be described
by a tiling model. We derived quite a number of tilings envolving fR,
sR, T, R, Q, K, and S tiles, all of which are compatible with a
decoration with atoms that include the original model. Although the
fR/Q and T/R models seem to be superior to the Q/K/S tiling and the
variants involving sR tiles, there may be properties that favour the
latter. We tested one variant (Q/K/S) of the tiling models again with
computer simulation to derive structure and potential parameters, and
to find out if the geometically reasonable model is also reasonable with a
simple interaction model. Since the Q/K/S tiling is somehow
cumbersome, but on the other hand includes most of the properties of
the other tilings, its stability assures us that the other tiling
types will also be stable. 

Up to now we have only treated each tiling for itself and described
its properties. In this final section we would like to address the
interrelationship of the tiling types, and if there is a hierarchy
of tilings. This includes the A/B layer symmetry breaking
which occurs in some of the models. In addition to computer
simulations it will be interesting to compare the tiling models to
experiment and to other structure models.
Furthermore we want to ask if and how it is possible to change
a certain tiling which means reshuffling the tiles, introducing
defects like random stacking and atomic jumps. These moves may also
lead to the a transformation of one tiling type into another. 
Although such a
process is not suitable for MD simulations, it may be
studied in Monte Carlo simulations with properly chosen interactions.

\subsection{ Hierarchy of Tiling Descriptions}

Our approach is an example of a general approach which may be fruitful
in understanding the relation of interatomic interactions to long-range
order in quasicrystals. An ensemble of tilings is proposed; 
via a decoration rule, these correspond one-to-one to an ensemble
of low-energy atomic structures. However, the energy 
is slightly different for each of these structures; these energy
differences may be considered as a ``tiling Hamiltonian'' 
${\cal H}_{tile}$
which is
purely a function of the tile arrangement. Often the tiling
Hamiltonian is a sum 
of one-tile and neighbor-tile pair interactions (Mihalcovi{\v c} et al.\
1996a,1996b) 
\nocite{mihI,mihII};
alternatively, Jeong and Steinhardt (1994) proposed a
family of tiling Hamiltonians with a (favorable) energy $-V$ for each
occurrence of a special local  pattern, which they called the ``cluster''.

The ground states of ${\cal H}_{tile}$ 
are a subset of the original ensemble; 
most often this constrained sub-ensemble can be described by a
tiling of larger (``super'') tiles with their own packing rules.
In the ``cluster'' Hamiltonians studied by Jeong and Steinhardt (1994), 
the degree of long-range order is enhanced in the supertiling --
indeed for one special ``cluster'' 
one just obtains the quasiperiodic  Penrose tiling --
and even a random tiling of supertiles has greatly reduced phason fluctuations.

Evidently, if 
${\cal H}_{tile}$
can be divided into 
a sum of successively weaker parts, there may be
a corresponding hierarchy of successively larger supertiles.
Furthermore, as the temperature is raised, the constraints 
due to the weakest terms of
${\cal H}_{tile}$
are broken, so the size of the relevant
tiles may decrease with temperature. (The associated changes in
diffraction pattern were discussed by Lan\c{c}on, Billard, Burkov \& de
Boissieu (1994)).
\nocite{burkov}
In Sec.~\ref{dsdm} , we discovered just such a hierarchy of similar
structures, some more constrained than others, as follows: the fR/Q tiling is
a special case of either the fR/sR/Q (fat and skinny rhombus and Q) tiling, or
of the T/R tiling. 
The 2-level (Q/K/S) tiling is also a special case of the fR/sR/Q, which is
a modification of the fR/sR tiling (Subsec.~\ref{basic}).

We postulated (at the start of Sec.~\ref{dsdm}) that
the dominant term in our tiling-Hamiltonian 
favors the chain-motif, consisting of $\alpha$ atoms surrounded by 
alternating pentagons. (Incidentally, this is a ``cluster''
in the sense of Jeong and Steinhardt  (1994).)
Then the sub-ensemble which minimizes that term is the T/R
random tiling, which (see below) is also consistent with 
a structure model proposed by He et al.\ \citeyear{chinese}.

Due to the limitations of our simulation, and because we did not 
perform relaxation  at $T=0$ to extract the parameters  of 
${\cal H}_{tile}$, 
we could not prove that the chain-motif (and hence T/R tiling)
is favored, nor could we
measure the smaller energy differences
that would decide which subensemble of T/R tilings 
most faithfully models the structure.
We could only conjecture (in Sec.~\ref{trirect})
that there should be an additional term 
in ${\cal H}_{tile}$, 
disfavoring Q tiles adjoining by $a$ edges. 
Granting that, it followed that
the best tiling to represent our structures is something
between the T/R tiling and the fR/Q tiling 
of Cockayne (1994).
\nocite{cock}

\subsection{ Issues of Symmetry Breaking in our
Simulation}

There remains a serious doubt that the apparent decagonal structure 
we observed in our simulations might be only an artifact of 
the geometric restrictions imposed by the small simulation box:
the correct phase in the thermodynamic limit
might be an icosahedral random tiling. 
Indeed, Sec.~\ref{icorel},
showed that our decagonal structure is essentially 
a special layered packing of the icosahedral tiles; 
rhombohedral tiling might 
find a layered arrangement in  a quench, if that happens
to fit well with the periodic boundary conditions.
Furthermore, out of seven quenches with 
the same potentials (Roth et al.\ 1995), \nocite{rotcool} 
five quenches led to icosahedral structures and two to decagonal ones.

Even if we grant that the true structure {\it is} decagonal, 
there is a second question whether we may be over-interpreting accidental
effects of the finite-size box used in our initial simulations:
does an A/B symmetry-breaking  occur in the thermodynamic limit?
This is the fundamental difference between the 2LT (Q/K/S) model
and the T/R tiling model.
Apart from the observation in our initial simulation, there is
no convincing reason for the 2LT (Q/K/S) model, since the K tile
seems to be poorly packed.
There is no other numerical evidence that the A/B symmetry-breaking
really occurs in an extended structure. 
It is quite possible that the shape and size of each layer
in our simiulation box happened to allow 
only a tiling which lacks the $180^\circ$ symmetry
axis (out of the plane) which would make A and B edges equivalent.\footnote{
A simple matching rule has been given that favours the
Q/K/S tiling over T/R and fR/Q: if you consider the pentagonal antiprisms
around the $\alpha$ atoms, then there should always be two different atoms on
the opposite sides of a diameter (true in the Q/K/S case). In the T/R tiling
there are only two vertex environments that fullfil this condition, in the
fR/Q tiling, there is only one. But with this vertex environments alone it is
not 
posssible to produce a fR/Q or T/R tiling. Thus the Q/K/S tiling is
favoured. The rule also works if you have sR tiles: In an sR/fR and an sR/fR/Q
tiling it either forces the assembly of Q/K/S tiles or induces chain shifts. 
}

To pursue this further, we should consider whether there 
is any physical cause for such a symmetry breaking.
Rather than start with the Q/K/S tiles, it may be better to start
by imagining a fR/sR/Q tiling and asking 
what might drive the symmetry breaking, 
which from this viewpoint is a displacive instability
of the $\gamma$ atoms, perhaps coupled to some chain-shifts. 
(After all, on every skinny rhombus such a symmetry-breaking occurs
locally; the question  is whether the pattern of displacements 
can be propagated from one skinny to the next.)
It is plausible that, once A/B symmetry breaking is granted, 
it will be advantageous to group the rhombi into S and K tiles. 

Both of these questions might be addressed (at zero temperature) by
relaxing a variety of decagonal and icosahedral structures under
a variety of potentials, to see which is more stable energetically,
in the spirit of calculations done on some Al-transition metal
quasicrystals (Phillips and Widom, 1994; Mihalcovi{\v c} et al.\ 1996b)
\nocite{phil,mihII}.

\subsection{ Comparison to Other Structure Models}
\label{comparison}

The Fe-Nb structure model outlined by He et al.\ \citeyear{chinese}
was not formulated in terms of a tiling of any kind, nor as
a deterministic rule for packing the plane with motifs;
instead, it is largely a scheme for analyzing high-resolution images.
Although there is a possibility that they misinterpreted images of 
a periodic approximant of the quasicrystal, 
or a microcrystalline mixture of approximants, 
it is still interesting and economical to decribe it using
quasicrystal tiling element.

Their model is based on the same motifs (pentagon-chains) as ours
and the lines in figure 3(b) of He et al.\ \citeyear{chinese},
which indeed
outline triangle and rectangle tiles, are the same as the $a$ linkages 
in our model. 
Since the sample they imaged
was rather small and defective (much like out first simulation), 
we can only be tentative
which variant of our structure model it should be identified with.
In principle a T/R tiling could be a
fR/Q tiling with the $b$ edges drawn in.
However, the image of He et al.\ \citeyear{chinese}
includes pairs of rectangles adjoining
by a $b$ edge, which is a defect from the fR/Q viewpoint;
note also it clearly does {\it not} have a A/B layer ordering.
Thus it is most plausibly idealized as a random T/R tiling; 

The icosahedral relationship described in Sec.~\ref{icorel} 
allows the decagonal structure to grow on the icosahedral
one. Congruent icosahedral/decagonal grain boundaries should therefore be
possible. The decagonal phase could be viewed as an approximant to the
icosahedral phase, and thin decagonal bands in the icosahedral phase may be
regarded as stacking defects. Furthermore, 
since the icosahedral and the
decagonal phases are similar in composition, 
phase transformations between them should also be
possible.
Thus our Lennard-Jones system might serve as a toy system for 
investigating the behavior $i$-AlPdMn.
(However, we must re-emphasize that
the atomic arrangement in an Al-transition metal 
quasicrystal certainly differs from the Frank-Kasper quasicrystal
described here.)

We turn briefly to another approach to constructing structure models --
that based on atomic clusters (Elser and Henley, 1985; Henley and Elser, 1986;
Mihalcovi{\v c} et al.\ 1996a)\nocite{elshen,henels,mihI},
not to be confused with the ``clusters'' of Jeong and Steinhardt (1994)!
Now, Al-Mn type structure models have a common motif of 
$\rm Al_6 Mn_4$ tetrahedra \cite{kreiner}; 
by joining several of these  along their faces larger clusters
are formed which are also observed in these structures. 
On the other hand, Frank-Kasper models have a common motif of
``truncated tetrahedra'' surrounding every L atom with coordination 16;
indeed, this motif is frequent in our structures. 
(see Fig.~\ref{friauf}; the full coordination shell, with 4 additional
atoms, is the ``Friauf polyhedron''.)
These are combined into larger clusters \cite{sam1,sam2} in exactly the same
fashion. 
Thus, if an Al-Mn model can be represented as packing of $\rm Al_6 Mn_4$
tetrahedra (plus atoms needed to fill the interstices), then by
replacing each of these by a truncated tetrahedron we produce a
(hypothetical) Frank-Kasper  model.
For example, the ``Mackay Icosahedron'' \cite{elshen} is 
a combination of 20 tetrahedra which maps to the
Bergman polyhedron (Bergman et al.\ 1957)\nocite{berg}.
It is well known that that crystal phases $\alpha$-AlMnSi and
$R$-AlCuLi are bcc packings of the respective clusters, 
and this suggested  that the same cluster-cluster networks could
describe the related icosahedral quasicrystals i-AlMnSi and i-AlCuLi
\cite{henCCT}.
A combination of 5 tetrahedra produces a pentagonal-bipyramid motif
which is the basis for crystalline approximants 
[such as $\rm  Al_{13}Fe_4$ and $\rm T_3(AlMnZn)$]
and for conjectured structure models of {\it decagonal} Al-transition metal
quasicrystals \cite{hendec}.
The corresponding combination in a Frank-Kasper structure 
[consisting of 5 truncated tetrahedra arranged around 
a central axis]
is the ``VF'' cluster, a common motif in large-unit-cell Frank-Kasper
alloys of simple metals \cite{sam1,sam2}. 
If we assume that such clusters are linked as in $\rm Al_{13}Fe_4$, 
we produce a new hypothetical Frank-Kasper decagonal model 
{\it different} from the one presented in the present paper. 
(The ``VF'' motif is rarer in our models -- it occurs only
when five fat rhombi meet to form a star.)
In the new structure model, the centers of neighboring clusters 
are separated by $\tau a$ in the horizontal planes
and by $\pm c/2$ vertically. We have not observed or investigated
such a model in simulations.

\subsection{ Beyond Two Dimensions}

To describe real structures, we must go beyond 
static, perfectly stacked structures. 
The real structure has 2 extra dimensions:
time, and the periodic direction $z$.
Whether its ultimate state is random-tiling or a locked quasiperiodic
tiling, a tiling can improve its order under annealing
only by reshuffling  of its tiles. 
For example, recoognizing the fundamental
reshuffling in the case of the binary tiling 
permitted an accelerated Monte Carlo move (Widom et al.\ 1987)\nocite{wid87}).
Note that, although the tile rearrangement apparently
involves a large volume, it is common that
the reshuffling requires the motion of only a single atom \cite{wid92}.

The problem of reshuffling has an obvious relation
to that of stacking disorder: a realistic model
must handle configurations in which the structure does not
repeat precisely along the fivefold axis.
Indeed, the structure as quenched
was not  strictly periodic in the stacking direction.
Of course, such nonperiodicity may be considered a defect, and
perhaps blamed on incomplete equilibration; however there are
two senses in which it is necessary even in equilibrium.

First, reaching equilibrium usually demands some tile reshufflings.
But in a 3D atomic structure based on a 2D tiling, every
tile reshuffling requires rearranging an entire column of the atomic structure.
Obviously this is easiest done one bit of the column at a time; the
intermediate state is one with stacking randomness.
Second, in the random-tiling  explanation of the thermodynamic
stability of quasicrystals, the contribution of tile-reshuffling 
entropy to the free energy is decisive; this entropy can be
extensive only in an ensemble with stacking disorder \cite{henART}.
Recently, Ritsch, Nissen, and Beeli \citeyear{ritsch} have argued that 
common features 
in high-resolution transmission electron microscope images of d(AlCoNi) 
are evidence of stacking randomness in this decagonal.\\[1ex]

\subsubsection{ Reshuffling}\label{reshuffling}

The reshufflings in the case of the two-level tiling 
are simple and are shown in Fig.~\ref{fig-old2}.
We can trade off $\rm Q+S \leftrightarrow 2 K$ 
(Fig.~\ref{fig-old2}(a))
or $\rm Q+K \leftrightarrow K+Q$ (Fig.~\ref{fig-old2}(b)), 
This reshuffling
is the simplest fundamental move which can be used to visit
from one state to another. 

Local reshufflings of the T/R tiling are not
possible; instead it is necessary to form a ``defect tile''
such as a skinny rhombus that is normally absent in that tiling.
A defect tile
can move through the tiling in what is called a ``zipper move''
\cite {ox} since it leaves behind a rearranged structure. 

One must check what reshufflings imply for
the atoms. In particular, the number of each species ought to be conserved;
otherwise the reshuffling is either blocked, 
or it creates substitutional defects, 
or it only takes place in  a structure which is already 
substitutionally disordered in equilibrium.

In a rhombus tiling, the phason strain 
(which can be regarded as the deviation
of the projection plane embedded in higher dimensional space from its ideal
position)
determines the number density of each kind of rhombus; since the rhombus 
content is conserved under reshuffling, so is the atomc content.
The same thing  is true in our T/R tiling, 
but not in the  two-level (Q/K/S) tiling, because of the 
possibility of trading off $\rm Q+S \leftrightarrow 2 K$. 
(When divided into Penrose rhombi, both combinations have
the same contents). 

Indeed,  our 2LT (Q/K/S) atomic structure model has serious problems with 
the reshufflings:
the $\rm Q+S \leftrightarrow 2 K$ reshuffling
changes the content
of the tiles from 29 L+ 14 S to 28 L +14 S, so one large atom disappears or
has to become an interstitial. The $\rm Q+K \leftrightarrow K+Q$ reshuffling
conserves the net atomic content of the tiles; 
however, the atomic rearrangement in which atoms jump the 
least distance would put atoms on the wrong site for their species.
Simulations have shown that such substitutions are not
acceptable; even a few of them destroy the stability of the structure.
We have to admit that no better moves have been found up to now. 

In a T/R tiling, the question of rearrangements is quite different
since the only possible update move is a ``zipper'', is an 
entire chain of tiles which closes on itself \cite{ox}, 
just as in triangle-square tilings \cite{oxhen}.
Thus, the intermediate state of a rearrangement involves not
only stacking defects, but also defects within each layer
(which would appear as special tiles other than T or R).

\subsubsection{ Stacking Randomness}

In a nonperiodic stacking, we may presume the
tilings describing adjacent layers are similar.
Thus the spatial sequence of layers is much like the
temporal sequence of a 2D tiling undergoing a series of reshufflings
\cite{henART}. 
Each violation of periodicity presumably costs energy;
most likely, one or two ways of doing so are less costly than any others,
so it is a reasonable approximation to postulate as 
a {\it constraint } on the stacked tiling ensemble that layers can 
be related spatially only in such ways (point stacking defects). 
In the case of the Q/K/S tiling model, the obvious stacking defects are 
flips $\rm Q+S \leftrightarrow 2 K$ and $\rm Q+K \leftrightarrow K+Q$
from one layer to the next. Since the positions of
the atoms are similar in both states, these defects should be not too costly in
energy. 

An interesting experiment would be
to create a stacking of layers from
{\it completely different} 2D tilings, and
see what structure it relaxes to. 
For the {\it dodecagonal} Frank-Kasper structure, 
in the monatomic case \cite{rothdiff},  
the atoms rearrange -- while moving only short distances -- until
all the layers are {\it identical} and defect free; on the other hand, 
in the binary case of the same dodecagonal structure 
(L and S atoms, similar to the present paper)
the structure cannot be healed with just
short-distance moves; instead substitutional defects are introduced.

\section*{ Acknowledgments}

J.R. would like to acknowledge a post doctoral fellowship from the DFG and the
hospitality at Cornell University, Ithaca, where most of this work has been
carried out. C.L.H. was supported by the U.S. D.O.E. grant DE-FG02ER89-45404.

\appendix
\section{Appendix: Inductive derivation of 2LT model}
\label{app-inductive}

Here we describe the details of an alternate inductive path to the
two-level tiling structure model of Sec.~\ref{twolevel}, using
the typical structural features that emerged from our simulation
(see Sec.~\ref{simulations}).
To start,  take the fundamental motif to be the chains of icosahedra;
let the quasilattice constant $a$ be the spacing between motif centers.
Adopt a zero-order description in which the 
pentagons in the chains are regular. 
In our simulation cell, the pentagons of these chains 
were linked by sharing edges in  the B layer, but by sharing corners 
in the A layer; this requires the B layer pentagons to be smaller
than the A layer pentagons.
Specifically the  pentagon radius (center to vertex) 
is $\tau^{-1} a$ in the B layer, and $a/2$ in the A layer. 

To zero order, the small A pentagons  should be S atoms
and the large B pentagons should be L atoms.
Placing these pentagons creates two kinds of too-close atom
pairs, to be resolved either by combining
two atoms into one or by relaxing them farther apart.

(i) In the B layer, whenever pentagon centers are separated by $b$, 
there will be a place (in a tile interior) 
where regular pentagon corners are separated by only $\tau^{-1}$ 
of a pentagon edge, too short to be a valid interatomic distance. 
This occurs whenever two motifs are second neighbors on the
tiling geometry, i.e. they are nearest neighbors 
to the same motif, forming an angle $2\pi/5$ with it.
These atoms relax perhaps $25\%$ further apart from each other. 
These are the $\gamma$ atoms and must become S atoms since they are
obviously being squeezed. 

(ii) In the A layer, the same pentagons whose centers are separated by $b$
also have corners very close to each other --
separated by $(\tau^{-2}/2)a \approx 0.19a$. 
In this case these sites should be merged, so that
these are the $\kappa, \lambda, \mu$ sites. Naturally, there is excess
space near these sites so they should become L atoms.

\vspace{0.5cm}

\bibliographystyle{newapa}\bibliography{deca4.bbl}

\newpage


\begin{table}[ht!]
  \label{table2}
  \begin{center}
    \begin{tabular}{c| c|c|c|c|c|c|c}
      layer & type & name & \# in Q & \# in K & \# in S & \# in 2T & \# in 2R
      \\ 
      \hline
      X&S & $\alpha$           & 4 & 6 & 10  & 2 & 4 \\
      \hline
      A&S & $\beta$            & 3 & 4 &  5  & 1 & 1\\
      A&L & $\kappa$           & 0 & 1 &  0  & 0 & 0 \\
      A&L & $\lambda$          & 0 & 0 &  5  & 1 & 0\\
      A&L & $\lambda^{\prime}$ & 2 & 3 &  0  & 0 & 2\\
      \hline
      B&S & $\gamma$           & 0 & 0 &  5  & 1 & 1\\
      B&S & $\gamma^{\prime}$  & 0 & 2 &  0  & 0 & 0\\
      B&S & $\delta$           & 2 & 2 &  0  & 0 & 2\\
      B&L & $\mu$              & 2 & 3 &  5  & 1 & 0\\
      B&L & $\mu^{\prime}$     & 0 & 0 &  0  & 0 & 2\\
      B&L & $\epsilon$         & 0 & 0 &  0  & 0 & 2\\
    \end{tabular}

    \caption{Content of the tiles. First column: layer. Second column: type of
      atom. The labels are introduced in
      Figs.~\ref{fig-basic}, \ref{qksdecl}. Third column: name of the
      site. Other columns: numbers in Q,K,S and T,R 
      tiles.} 
  \end {center}
\end{table}

\begin{table}[ht!]
  \label{table1}
  \begin{center}
    \begin{tabular}{c|c|c|ccc}
      layer & name & coord. & \multicolumn{3}{c}{frequency}\\
      \hline
      X & $\alpha$ & 12 & 2$(\tau+\tau^{-5})/\sqrt{5}$ & = & 1.52786\\ 
      \hline
      A & $\beta$ & 12 & 1 & = & 1.00000 \\ 
      A & $\kappa$ & 15* & $\tau^{-5}$ & = & 0.09017 \\ 
      A & $\lambda$ & 16 & $\sqrt{5}\tau^{-5}$ & = & 0.20163 \\
      A & $\lambda^{\prime}$ & 15 & $(2\tau+3)/\tau^{5}$ & = & 0.56231 \\
      \hline
      B & $\gamma$& 12 & $\sqrt{5}\tau^{-5}$ & = & 0.20163 \\
      B & $\gamma^{\prime}$ & 12* & 2$\tau^{-5}$ & = & 0.18034 \\
      B & $\delta$ & 14 & $2(\sqrt{5}-2)$ & = & 0.47214 \\
      B & $\mu$ & 16 & $(5+\sqrt{5})\tau^{-5}$ & = & 0.65254  \\
      \hline
      \hline
      X & $\alpha$ & 12 & 2$\tau$ & = & 3.23607\\
      \hline
      A & $\beta$ & 12 & (3$\tau$-2)/2 & = & 1.42705\\
      A & $\epsilon$ & 14 & $\tau^{-2}$ & = & 0.38197 \\ 
      A & $\lambda$ & 16 & $2\tau^{-1}$ & = & 1.23607 \\ 
      A & $\lambda^{\prime}$ & 15 & $\tau^{-2}$ & = & 0.38197\\
      \hline
      B & $\gamma$ & 12 & (3$\tau$-2)/2 & = & 1.42705\\
      B & $\delta$ & 14 & $\tau^{-2}$ & = & 0.38197 \\ 
      B & $\mu$ & 16 & $2\tau^{-1}$ & = & 1.23607 \\ 
      B & $\mu^{\prime}$ & 15 & $\tau^{-2}$ & = & 0.38197
    \end{tabular}
    
    \caption{ Coordination polyhedra for the two level tiling (Q/K/S) and the
      triangle/rectangle (T/R) tiling. The fR/Q tiling can be derived from the
      T/R tiling by a subdivision of the tiles as indicated in
      Fig.~\ref{fig-basic}. First column: layer (A,B,X). Second column: name
      of the site. The prime indicates a coordination number different from
      the standard. Third column: coordination number, the star indicates
      non-FK-coordination. Fourth column: relative frequency. The
      composition of the two level tiling is S:L = 0.692:0.382, 
      the composition of the T/R tiling is S:L = 0.679:0.321.} 
  \end{center}
\end{table}

\clearpage


\begin{figure}[ht!]
\centerline{\includegraphics[width=10cm]{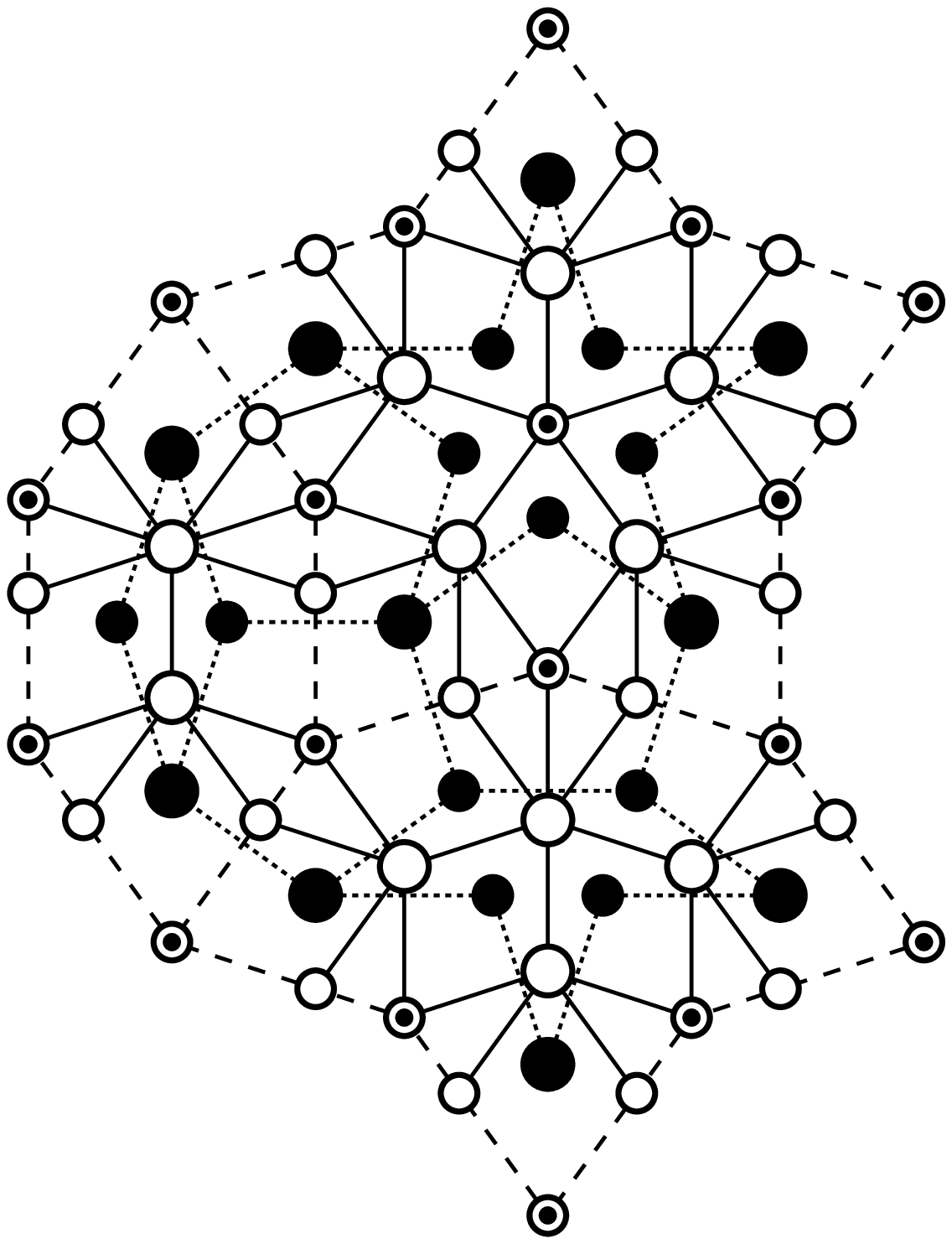}}
\caption{Auxiliary tiling for the acceptance domains, topologically equivalent
  to the Q/K/S decoration, adapted to have integer coordinates. If the atoms
  on the   large Penrose tile edges are shifted, the result is a deflated
  Penrose tiling but with variing decoration of the vertices. Long dashes:
  original tiles, short dashes: pentagon tiling, full and dashed lines indicate
  the small Penrose tiles. }
\label {fig-penrose} 
\end{figure}

\begin{figure}[ht!]
\centerline{\includegraphics[width=7cm]{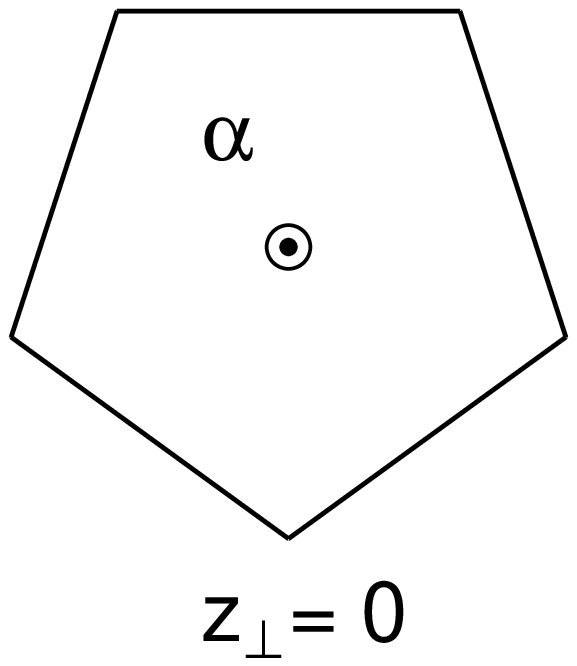}}
\centerline{\includegraphics[width=7cm]{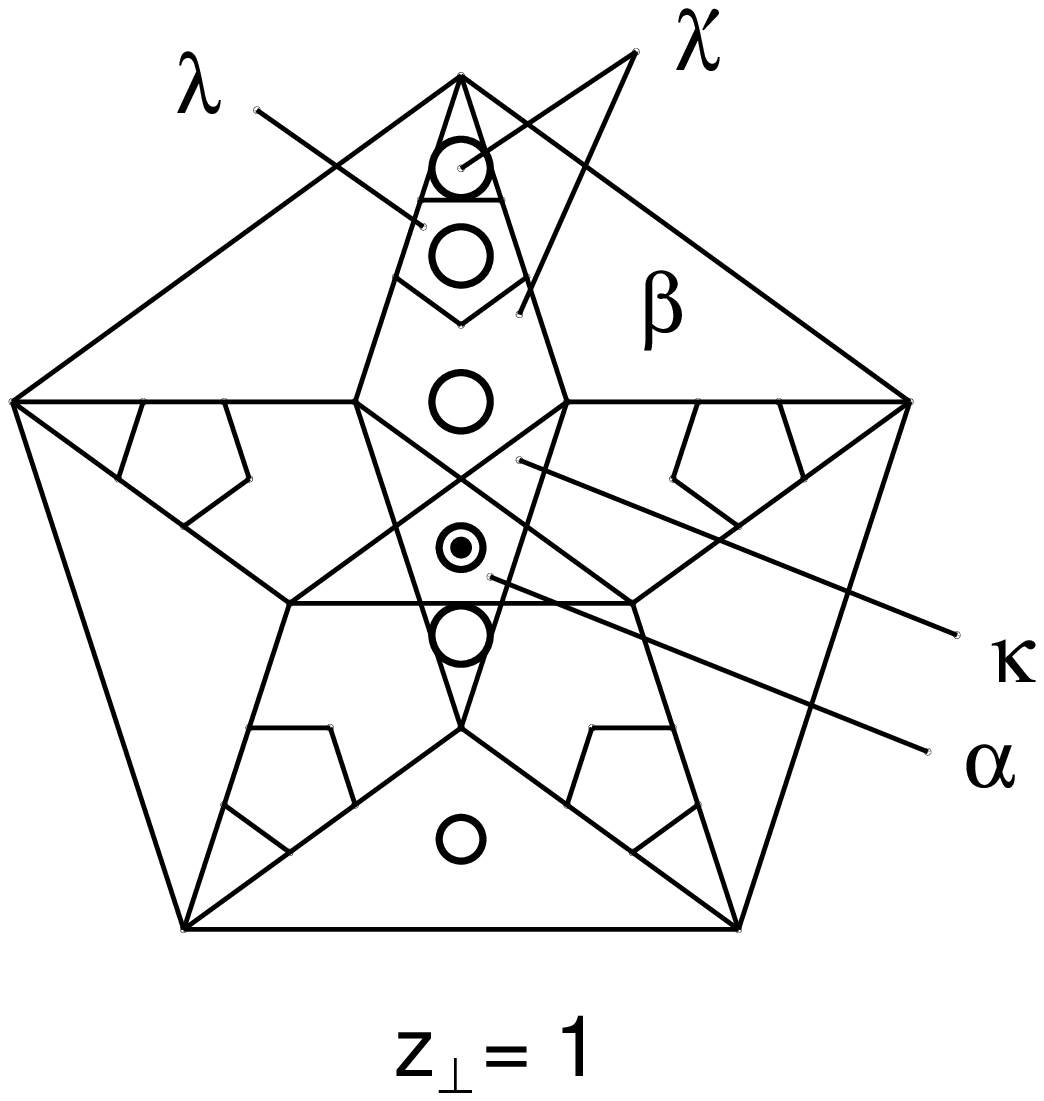}}
\centerline{\includegraphics[width=7cm]{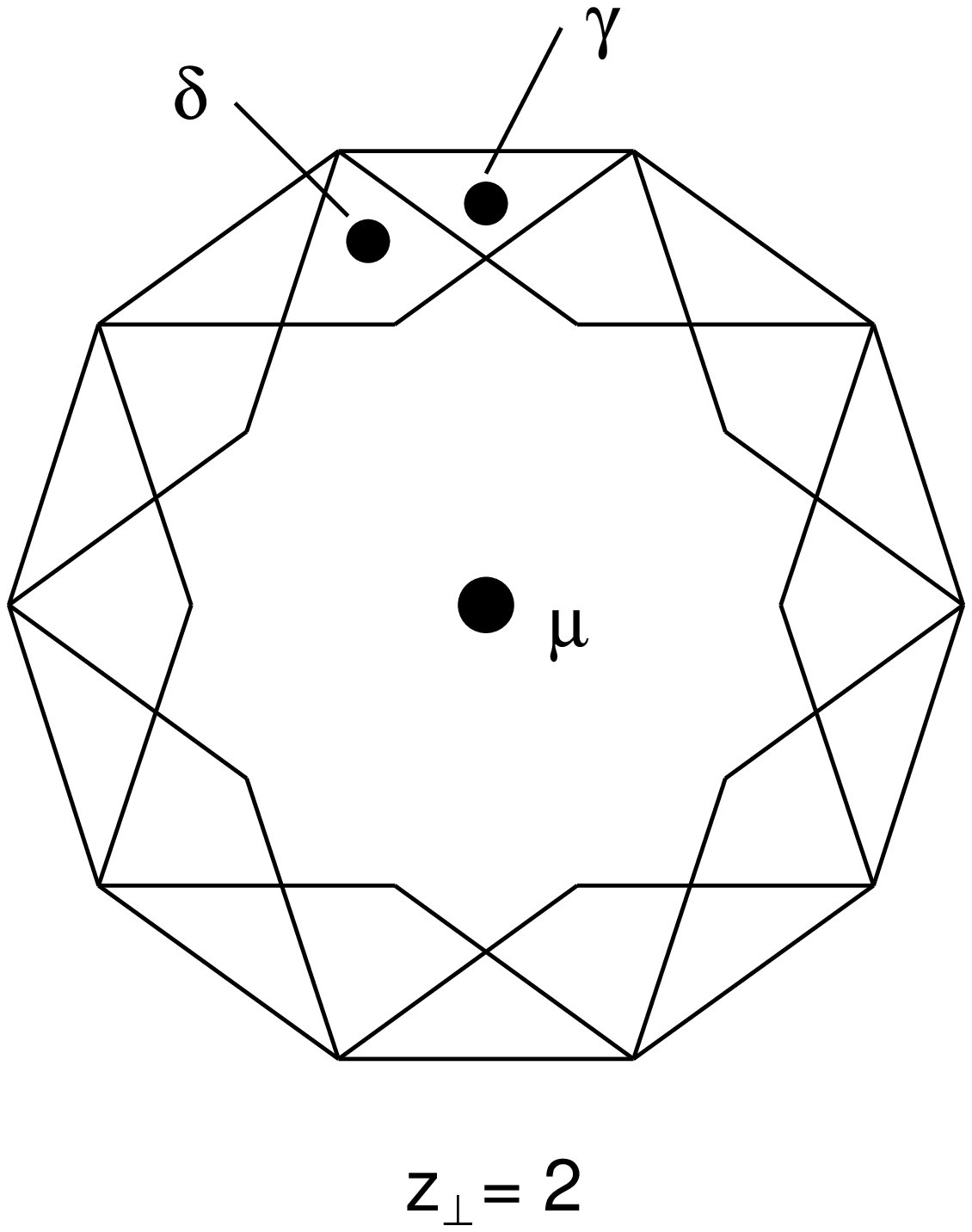}}
\caption{ Acceptance domain of the Q/K/S tiling.
Here ``$z_{\bot}$'' labels the commensurate component of perp-space.
Greek letters show the atom class; the layer and species are
indicated also. The symbols for the atoms are the same as in the other
figures.  
}
\label {fig-acc}
\end{figure}
 
\begin{figure}[ht!]
\centerline{\includegraphics[width=5cm]{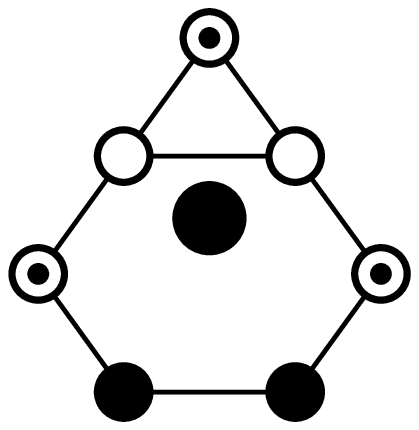}}

\caption{ Truncated tetrahedron (solid lines) with $\mu$ atom at 
center,  showing its 
placement in the fat Penrose rhombus decoration. (The rhombus contains
another truncated tetrahedron centered on a $\lambda$ atom, 
which is not shown.) 
}
\label {friauf}
\end{figure}
 
\begin{figure}[ht!]
\centerline{a)\includegraphics[width=10cm]{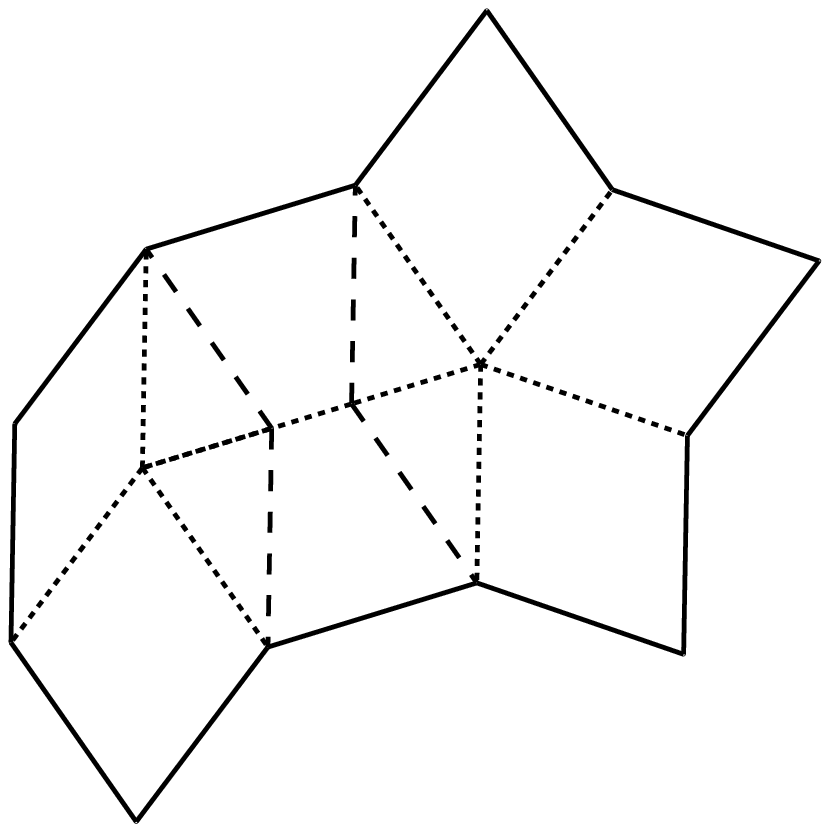}}
\centerline{b)\includegraphics[width=10cm]{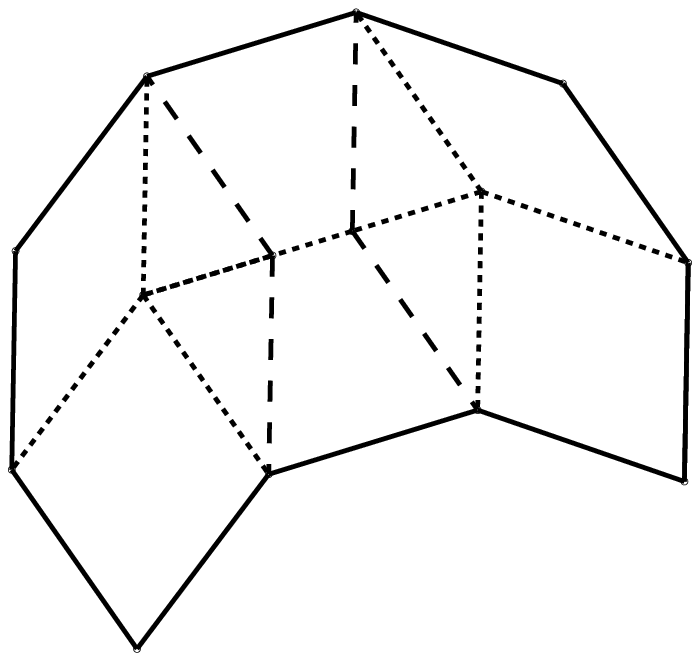}}
\caption{ 
Reshuffling operations in the two-level tiling.
(a)
$\rm Q+S \leftrightarrow 2 K$
(b)
$\rm Q+K \leftrightarrow K+Q$.
In each case, the dotted lines show the Penrose tiles into
which the Q, K, and S tiles may be subdivided;
both reshufflings correspond to the same reshuffling of the Penrose tiles.
}
\label {fig-old2}
\end{figure}

\begin{figure}[ht!]
\centerline{a)\includegraphics[width=10cm]{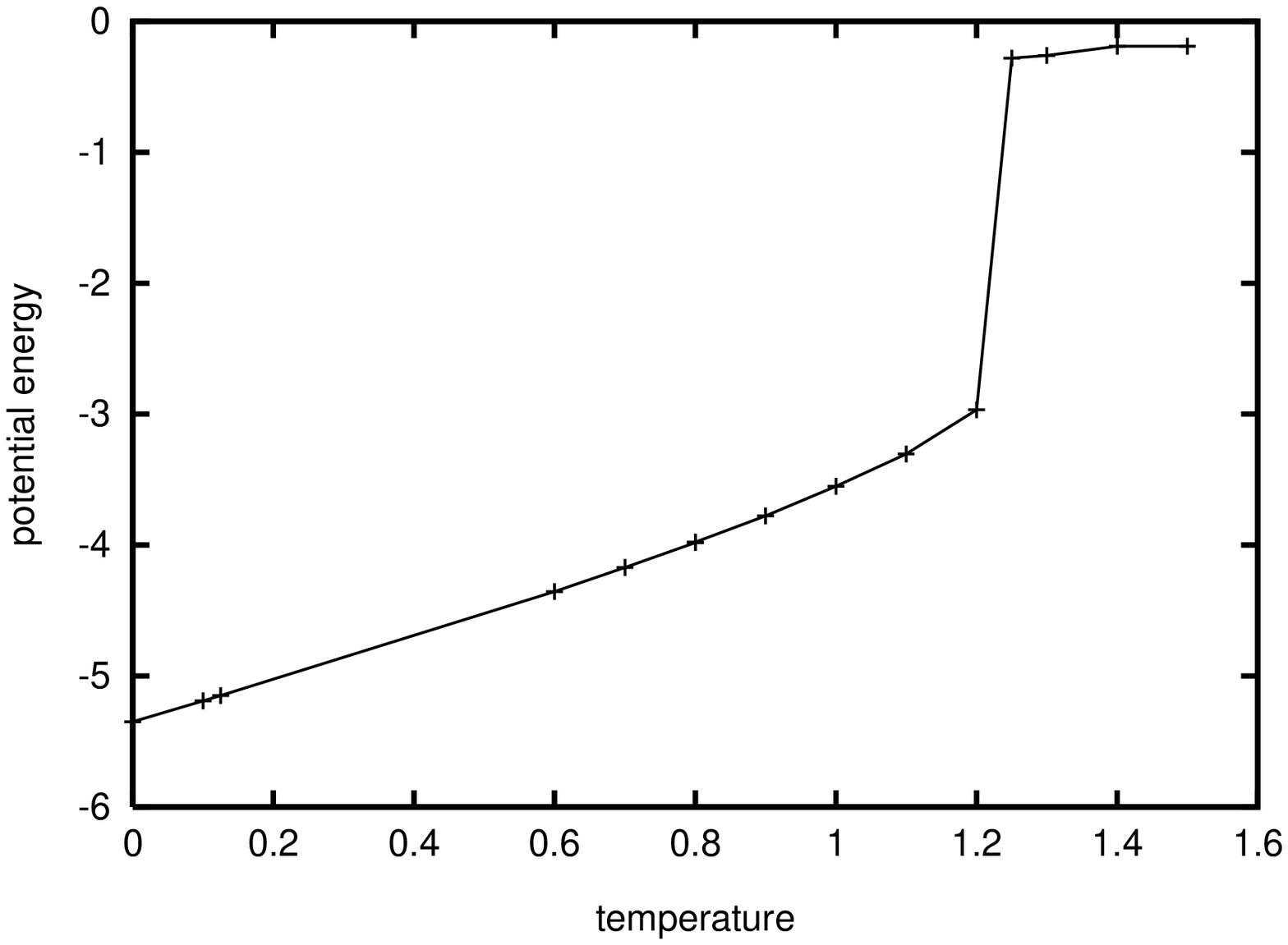}}
\centerline{b)\includegraphics[width=10cm]{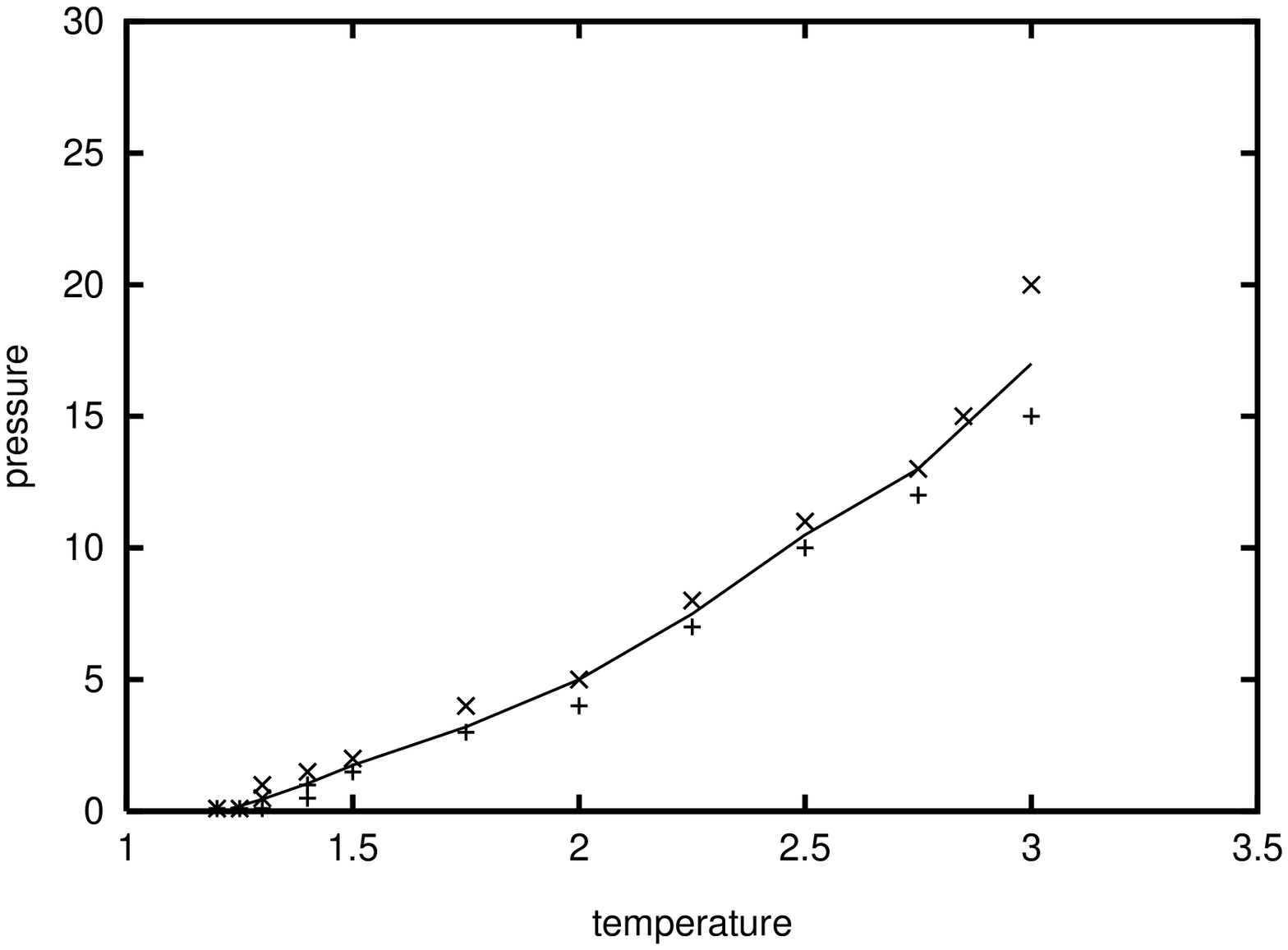}}
\caption{
Thermodynamic properties from simulations with Dzugutov's potentials.
(a) Potential energy {\it vs.} temperature at $P=0.1$ 
(b) Phase diagram for our binary decagonal quasicrystal.
The sublimation line is shown solid. The crosses and plus signs indicate the
actual simulation runs to determine the phase boundary.
Note the absence of a liquid phase!
}
\label{fig-thermo}
\end{figure}

\end{document}